\documentclass[letter,11pt]{article}
\pdfoutput=1
\usepackage{jheppub}
\usepackage{hyperref}
\usepackage{graphicx,epstopdf,amsmath,amsfonts,amssymb,appendix,comment} 
\usepackage{color,slashed,subfigure,setspace,footnote,multirow,longtable,braket}
\usepackage{epsfig,mathrsfs,latexsym,color,url,etoolbox}
\usepackage{bbm,dsfont}
\definecolor{nicered}{rgb}{0.7,0.1,0.1}
\definecolor{nicegreen}{rgb}{0.1,0.5,0.1}
\definecolor{CarnationPink}{rgb}{1.0, 0.65, 0.79}
\DeclareMathAlphabet{\mathbbold}{U}{bbold}{m}{n}    

\usepackage{comment}
\usepackage{tensor}
\usepackage{cases}
\usepackage{ulem}

\usepackage{float}

\allowdisplaybreaks

\usepackage{dcolumn}
\usepackage{bm}
\usepackage[table]{xcolor}
\usepackage{multirow}
\usepackage{comment}
\usepackage{url}
\usepackage{tcolorbox}
\usepackage[T1]{fontenc}

\newcommand{\absq}[1]{\left\lvert #1 \right\rvert^2}

\DeclareMathAlphabet{\mathpzc}{OT1}{pzc}{m}{it}

\newcommand{\eg}{{\it e.g.}}
\newcommand{\ie}{{\it i.e.}}

\newcommand{\bea}{\begin{align}}
\newcommand{\eea}{\end{align}}

\newcommand{\afb}{\ensuremath{A_{\mathrm{FB}}}}

\def\Fermilab{Theoretical Physics Department, Fermilab, P.O. Box 500, Batavia, IL 60510, USA}
\def\Northwestern{Department of Physics \& Astronomy, Northwestern University, Evanston, IL 60208, USA}

\begin{document}

\preprint{FERMILAB-PUB-21-180-T, NUHEP-TH/21-01}

\title{Three-Body Decays of Heavy Dirac and Majorana Fermions}

\author[1]{Andr{\'e} de Gouv{\^e}a,}
\author[2]{Patrick J. Fox,}
\author[2]{Boris J. Kayser,}
\author[2]{Kevin J. Kelly}
\affiliation[1]{\Northwestern}
\affiliation[2]{\Fermilab}

\emailAdd{degouvea@northwestern.edu}
\emailAdd{boris@fnal.gov}
\emailAdd{pjfox@fnal.gov}
\emailAdd{kkelly12@fnal.gov}

\date{\today}

\abstract{
Nonzero neutrino masses imply the existence of degrees of freedom and interactions beyond those in the Standard Model. A powerful indicator of what these might be is the nature of the massive neutrinos: Dirac fermions versus Majorana fermions. While addressing the nature of neutrinos is often associated with searches for lepton-number violation, there are several other features that distinguish Majorana from Dirac fermions. Here, we compute in great detail the kinematics of the daughters of the decays into charged-leptons and neutrinos of hypothetical heavy neutral leptons at rest. We allow for the decay to be mediated by the most general four-fermion interaction Lagrangian.  We demonstrate, for example, that when the daughter charged-leptons have the same flavor or the detector is insensitive to their charges, polarized Majorana-fermion decays have zero forward/backward asymmetry in the direction of the outgoing neutrino (relative to the parent spin), whereas Dirac-fermion decays can have large asymmetries. Going beyond studying forward/backward asymmetries, we also explore the fully-differential width of the three-body decays. It contains a wealth of information not only about the nature of the new fermions but also the nature of the interactions behind their decays.
}

\maketitle
\flushbottom

\clearpage
\newpage
\section{Introduction}
\label{sec:Introduction}

Massive fermions with no conserved quantum numbers can be either Majorana fermions or Dirac fermions. At present, there are no identified fundamental Majorana fermions in nature. A few decades ago, this statement was neither especially meaningful nor surprising. All known fundamental fermions are charged under unbroken gauge symmetries (e.g., the electromagnetic $U(1)$) except for neutrinos which, until the end of the twentieth century, could be considered exactly massless. With the discovery of nonzero neutrino masses, research into mechanisms to test the hypothesis that neutrinos are Majorana fermions has grown in volume and impact.\footnote{Concurrently, the evidence for dark matter has also grown very significant over the last decade. The hypothesis that dark matter is a new fundamental particle is very attractive and under intense experimental and theoretical scrutiny. Should this hypothesis be verified, and should the dark matter particle turn out to be a heavy, neutral fermion, determining its nature will also become an urgent question for particle physics.}

Nonzero neutrino masses also imply the existence of new degrees of freedom. Currently, their nature and properties are very poorly constrained. The new degrees of freedom associated to nonzero neutrino masses could be bosons or fermions, charged or neutral, very heavy or very light. One popular scenario postulates the existence of new massive Majorana fermions that mix with the Standard Model neutrinos. In the event of the discovery of a new neutral lepton -- a heavy neutrino or, as is more common in the literature, a heavy neutral lepton (HNL) -- identifying its nature -- Majorana fermion (MF) or Dirac fermion (DF) --  will become an urgent question for particle physics. HNLs, independent from their possible connection to the observed neutrino masses, are also a candidate for the dark matter and remain an ingredient of potential solutions to the so-called short-baseline anomalies. HNLs are the subject of experimental searches at all mass scales \cite{deGouvea:2015euy,Drewes:2015iva,Fernandez-Martinez:2016lgt,Drewes:2016jae,Bryman:2019ssi,Bryman:2019bjg,Bolton:2019pcu}.   

A very promising way to determine the nature of the neutrinos, including HNLs, is to test the hypothesis that global lepton number is conserved in nature. On the one hand, if lepton number is a symmetry of nature, massive neutrinos must be Dirac fermions since neutrinos are non-trivially charged under lepton number in such a way that the neutrino state and the antineutrino state are distinguishable. On the other hand, if lepton number is violated (by two units) then neutrinos are Majorana fermions. The deepest probes for the violation of lepton number are searches for the neutrinoless double-beta decay of various nuclei. These are the subject of intense experimental research (see \cite{Dolinski:2019nrj} for a recent review).

There are ways to distinguish Majorana from Dirac fermions that do not directly involve searches for the violation of some symmetry. Majorana fermions are their own antiparticles while Dirac fermions are not. Hence, processes where Dirac and Majorana fermions are created or destroyed are distinct and their measurable properties -- differential cross-sections and decay rates -- are, in principle, recognizably different. There are several identified examples of this, including (i) different velocity dependence (near threshold) for fermion--antifermion annihilation, a fact that has important consequences for dark matter phenomenology, (ii) the decay of a Majorana fermion into two self-conjugate final-state particles is, at leading order, isotropic in the rest frame of the parent independent from the physics responsible for the decay \cite{Balantekin:2018azf,Balantekin:2018ukw}, (iii) the rate of cosmic-background neutrino capture on tritium is twice as large if the neutrinos are Majorana fermions relative to Dirac fermions \cite{Long:2014zva}, (iv) Majorana fermions have zero electromagnetic moments (transition moments, however, are allowed), (v) there are large differences in the rates and kinematics of neutral-current decays of atoms, very low-energy electron-photon scattering \cite{Berryman:2018qxn}, etc.  

Here, we explore in detail the differential decay rate of polarized Majorana and Dirac fermions and how these compare to one another. As is the case of two-body decays, the allowed kinematical distributions of the daughter particles in three-body decays are more constrained if the parent particle is a Majorana fermion. This means that, in principle, there are circumstances under which, if the parent particle is a Dirac fermion, one can rule out the ``wrong'' hypothesis -- Majorana fermion in this case. The converse is not true unless one has independent information on the physics responsible for the three-body decay.  

We cast our discussion in the context of HNLs and explore their decays into standard model charged-leptons and neutrinos; nonetheless, many of the results discussed here should apply much more broadly. We consider that the decay is mediated by the most general four-fermion effective Lagrangian and hence our results are not constrained by the idiosyncrasies of the Standard Model weak interactions.

In section~\ref{sec:CPTArguments}, we present arguments based upon the CPT properties of the final states to show that if the HNL is a Majorana fermion in certain classes of decays there is no forward/backward asymmetry of the charged-lepton-pair (dilepton) system relative to the spin of the HNL.  In section~\ref{sec:DirMajAmp}, we discuss the most general possible matrix element for the decay of a polarized HNL, for both a Dirac or Majorana fermion.  In section~\ref{sec:Anisotropy}, we discuss the size of potential kinematic features in HNL three-body decays for various well motivated choices of couplings.  Much of the technical details of the calculations are relegated to appendices \ref{appendix:MSqs}-\ref{eq:TIntegrals}.
In section~\ref{sec:DiracFakingMajorana}, we discuss circumstances under which it is possible for observations to be consistent with both the MF and DF hypotheses. In section~\ref{sec:QuadDiff}, we extend our discussion beyond the forward/backward asymmetry of the dilepton system to include the full differential distributions and discuss how analyzing these distributions allows for further separation of the Dirac and Majorana hypotheses, as well as allowing for distinction between certain coupling structures.  We conclude in section~\ref{sec:Conclusions}.

\section{Forward/Backward Symmetry of Fermion Decays from CPT}
\label{sec:CPTArguments}

Refs.~\cite{Balantekin:2018azf,Balantekin:2018ukw} demonstrated that the two-body decay of a polarized Majorana fermion $N$ into a light Majorana neutrino $\nu$ and a self-conjugate boson $X^0$ is isotropic \ie\ the differential partial width $d\Gamma/d\Omega$ is constant.  
On the contrary, two-body decays of a Dirac fermion $N$ may have strong $\cos\theta_X$ dependence, where $\theta_X$ is the direction of the outgoing $X^0$ relative to the spin of $N$ in the rest frame of the decaying $N$.

General two-body decays $N \to \nu X^0$, when $N$ is 100\% polarized, can be expressed, in the $N$ rest frame, as
\begin{equation}
\label{eq:dGammadcostheta}
\frac{d\Gamma(N \to \nu X^0)}{d \cos{\theta_X}} = \frac{\Gamma}{2} \left(1 + 2A_{\rm FB} \cos\theta_X\right),
\end{equation}
where the forward/backward asymmetry $A_{\rm FB}$ is defined as
\begin{align}
\label{eq:FBasymmetry}
A_{\rm FB} \equiv \displaystyle\frac{\displaystyle\int_{0}^{1} \frac{d\Gamma}{d\cos\theta_X} d\cos\theta_X - \displaystyle\int_{-1}^{0} \frac{d\Gamma}{d\cos\theta_X} d\cos\theta_X}{\displaystyle\int_{0}^{1} \frac{d\Gamma}{d\cos\theta_X} d\cos\theta_X + \displaystyle\int_{-1}^{0} \frac{d\Gamma}{d\cos\theta_X} d\cos\theta_X}.
\end{align}
Assuming only CPT-invariance, Refs.~\cite{Balantekin:2018azf,Balantekin:2018ukw} demonstrated that, when $X$ is a self-conjugate boson, at leading order,\footnote{The result is exact if CP-symmetry is strictly conserved.} $A_{\rm FB}$ is zero when $N$ is a Majorana fermion. Indeed, the decays of Majorana $N$ are isotropic in terms of the direction of the outgoing $X$ (or equivalently the direction of the outgoing neutrino).

Extending the results of Refs.~\cite{Balantekin:2018azf,Balantekin:2018ukw} to three-body decays of MF and DF requires additional considerations because the final-state phase space now depends on \textit{\textbf{five}} kinematic variables\footnote{We define this phase space in Section~\ref{sec:Anisotropy} and discuss it further in Appendix~\ref{app:KinIntegration}.}. However, if it is possible to interpret two of the three final-state particles as a ``system'' $X^0$ (as above) with definite CPT properties, we can, based on these CPT properties, make connections between the two-body and three-body decays.

For clarity, we will focus on the decay of $N$ into a neutrino $\nu$ and two charged leptons $\ell_\alpha^-$ and $\ell_\beta^+$, $\alpha,\beta=e,\mu,\tau$.  When applying CPT arguments, we consider two separate cases -- one in which the final-state charged leptons have identical flavor ($\alpha = \beta$), and one in which their flavors are distinct but whatever detector is measuring these final-state particles cannot determine the charge of the individual particles on an event-by-event basis. For both of these situations, we will be considering the charged lepton system $\ell_\alpha^+ \ell_\beta^-$ as a single system, which we refer to as $X^0$ with a (variable)  invariant mass $m_{\ell\ell}^2$. This will allow us to express the decay $N \to \nu \ell_\alpha^+ \ell_\beta^-$ as a pseudo-two-body decay $N \to \nu X^0$. 

In order to consider the final-state particles $\ell_\alpha^- \ell_\beta^+$ as a system $X^0$, we must integrate over the kinematical quantities in the three-body phase space that contain internal information regarding the individual four-momenta of $\ell_\alpha^-$ and $\ell_\beta^+$. Thus, the CPT arguments \textit{will not} be able to determine whether the decays $N \to \nu \ell_\alpha^- \ell_\beta^+$ are (an)isotropic, but they \textit{will} allow us to determine that the MF decays are isotropic in terms of the direction of the system $X^0$.

\subsection{Same Flavor Final-State Charged Leptons}
\label{subsec:CPTIdentical}

In the case of same flavor final state leptons, we define $X^0 \equiv \ell_\alpha^+ \ell_\alpha^-$ and $X^0$ is self-conjugate. Additionally, we define $\lambda_\nu$ and $\lambda_X$ to be the spin-projection of the outgoing neutrino and $X^0$ along their respective directions of motion and $\lambda \equiv \lambda_\nu - \lambda_X$. 
Here, and throughout, we assume that the operators generating the decay of $N$ are of the four-fermion type, i.e. $N$ can be thought of as a point-like particle decaying through a contact interaction.
Unless otherwise noted, we are in the reference frame where $N$ is at rest and, in these calculations, we assume $N$ and $\nu$ are Majorana fermions and that $N$ is polarized in the spin-up direction, denoted by ``$\uparrow$''. Defining $\Gamma_{\lambda=+1/2}$ and $\Gamma_{\lambda=-1/2}$ to be the corresponding partial widths for decays with $\lambda = +1/2$ and $\lambda=-1/2$, respectively, we may express the decay $N \to \nu X^0$ as
\begin{equation}\label{eq:TwoBodyMaj}
\frac{d\Gamma (N \to \nu X^0)}{d\cos\theta_{X}} = \frac{1}{2}\Gamma_{\lambda=+1/2} \left(1 + \cos\theta_X\right) + \frac{1}{2}\Gamma_{\lambda=-1/2} \left(1 - \cos\theta_X\right).
\end{equation}
If we define the momenta of $\nu$ and $X^0$ to be $\vec{q}$ and $-\vec{q}$, respectively, we can express the leading-order transition-amplitude-squared for the spin-up decay as
\begin{equation}
\left\lvert \mathcal{A}\right\rvert^2 = \left\lvert \braket{\nu(\vec{q},\lambda_\nu) X^0 (-\vec{q}, \lambda_X)| H | N(\uparrow)}\right\rvert^2,\label{eq:Asq0}
\end{equation}
where $H$ is the interaction Hamiltonian governing this decay. If we apply CPT to $\left\lvert \mathcal{A}\right\rvert^2$, defining the operator $\xi$ as the action of CPT, and if we assume that the Hamiltonian is CPT-invariant, we obtain
\begin{align}
\left\lvert \mathcal{A}\right\rvert^2 &= \left\lvert \braket{ \xi H \xi^{-1} \xi N(\uparrow) | \xi \nu(\vec{q},\lambda_\nu) X^0(-\vec{q},\lambda_X)}\right\rvert^2, \\
&= \left\lvert \braket{ \nu(\vec{q},-\lambda_\nu) X^0(-\vec{q},-\lambda_X) | H | N(\downarrow)}\right\rvert^2, \\
&= \left\lvert \braket{\nu(-\vec{q},-\lambda_\nu) X^0(\vec{q},-\lambda_X) | H | N(\uparrow)}\right\rvert^2,\label{eq:Asq1}
\end{align}
where the last line is obtained by rotation of the system by an angle $\pi$ about an axis perpendicular to both the $N$ spin-direction and $\vec{q}$. Summing expressions~\eqref{eq:Asq0} and~\eqref{eq:Asq1} over the helicities for which $\lambda_\nu - \lambda_X \equiv \lambda = +1/2$, and comparing~\eqref{eq:Asq0} to~\eqref{eq:Asq1}, we see that $\Gamma_{\lambda=+1/2} = \Gamma_{\lambda=-1/2}$. From Eq.~\eqref{eq:TwoBodyMaj}, $d\Gamma/d\cos\theta_X = \Gamma/2$, a constant, and this implies the $X^0$ direction distribution is isotropic. 

\subsection{Charge-Blind Detector}
\label{subsec:CPTChargeBlind}
Here, we define $X^0 \equiv \ell_\alpha^+ \ell_\beta^-$ ($\alpha\neq\beta$) and note that now, $X^0 \neq \overline{X^0}$: it is \textit{not} a self-conjugate state. However, we assume that our detector is charge blind and cannot distinguish between these two states, and so the object we are interested in is the sum of two differential partial widths $N \to \nu X^0$ and $N \to \nu \overline{X^0}$. The differential width for $N \to \nu X^0$ follows the same form as Eq.~\eqref{eq:TwoBodyMaj}, while the decay $N \to \nu \overline{X^0}$ takes the form
\begin{equation}
\frac{d\Gamma(N \to \nu \overline{X^0})}{d\cos\theta_X} = \frac{1}{2} \overline{\Gamma}_{\lambda=+1/2} (1 + \cos\theta_X) + \frac{1}{2}\overline{\Gamma}_{\lambda=-1/2} (1 - \cos\theta_X),
\end{equation}
where $\overline{\Gamma}_{\lambda = \pm 1/2}$ refer to the partial widths of these decays for $\lambda = \pm 1/2$. Using the same CPT application as above, we can now relate the decays by
\begin{align}
\left\lvert \mathcal{A}\right\rvert^2 &= \left\lvert \braket{\nu(\vec{q},\lambda_\nu) X^0(-\vec{q},\lambda_X) | H | N(\uparrow)}\right\rvert^2, \\
&= \left \lvert \braket{ \xi H \xi^{-1} \xi N(\uparrow) | \xi \nu(\vec{q},\lambda_\nu) X^0 (-\vec{q},\lambda_X)} \right \vert^2, \\
&= \left\lvert \braket{ \nu(\vec{q}, -\lambda_\nu) \overline{X^0}(-\vec{q},-\lambda_X) | H | N(\downarrow)}\right\rvert^2, \\
&= \left\lvert \braket{\nu(-\vec{q}, -\lambda_\nu) \overline{X^0}(\vec{q}, -\lambda_X) | H | N(\uparrow)}\right\rvert^2.
\end{align}
Again, the last line is obtained by a $\pi$-rotation around an axis perpendicular to the decay plane.  Here, because $X^0$ is no longer self-conjugate we cannot relate $\Gamma_{\lambda=+1/2}$ to $\Gamma_{\lambda=-1/2}$. Instead, we obtain (after summing over the unobserved $\lambda_\nu$ and $\lambda_X$) $\Gamma_{\lambda=+1/2} = \overline{\Gamma}_{\lambda=-1/2}$ and $\Gamma_{\lambda=-1/2} = \overline{\Gamma}_{\lambda=1/2}$. Then, the object we wish to calculate is
\begin{align}
\frac{d\Gamma}{d\cos\theta_X} &\equiv \frac{d\Gamma(N\to \nu X^0)}{d\cos\theta_X} + \frac{d\Gamma(N\to\nu \overline{X^0})}{d\cos\theta_X}, \\
&= \frac{1}{2} \Gamma_{\lambda=+1/2} (1+\cos\theta_X) + \frac{1}{2}\Gamma_{\lambda=-1/2}(1 - \cos\theta_X) + \frac{1}{2} \Gamma_{\lambda=+1/2} (1- \cos\theta_X) + \frac{1}{2}\Gamma_{\lambda=-1/2}(1 + \cos\theta_X), \\
&= \Gamma_{\lambda = +1/2} + \Gamma_{\lambda = -1/2}.
\end{align}
Again, we see a flat distribution with respect to $\cos\theta_X$ -- this will yield zero forward-backward asymmetry.

\section{General Amplitudes for Heavy Neutrino Decay}
\label{sec:DirMajAmp}
\setcounter{footnote}{0}
We now consider the entire three-body phase space of MF and DF decays.
Many studies of heavy neutral leptons and their decays exist in the literature~\cite{Formaggio:1998zn,Gorbunov:2007ak,Asaka:2012bb,Ballett:2016opr,Berryman:2017twh,Coloma:2017ppo,Arbelaez:2017zqq,Cvetic:2018elt,Bondarenko:2018ptm,Curtin:2018mvb,SHiP:2018xqw,Ariga:2018uku,Krasnov:2019kdc,Abe:2019kgx,Ballett:2019bgd,Drewes:2019byd,Chun:2019nwi,Arguelles:2019ziu,Abratenko:2019kez,Berryman:2019dme,Gorbunov:2020rjx,Coloma:2020lgy,Batell:2020vqn,deVries:2020qns,Plestid:2020ssy,Breitbach:2021gvv}, but the focus of these is usually on the scenario in which the new fermion's only interactions with the Standard Model (SM) are via mixing with the light, SM neutrinos. This predicts that its decays are mediated by the SM $W$ and $Z$ bosons, and the interaction structure of the decays is known.

Here, we define a general framework for describing the decays of MF and DF in a way that is mostly independent from the nature of the new-physics interactions. We focus on the scenario where the new particle decays to a SM neutrino and a pair of charged leptons. However, this framework can apply for decays into three light neutrinos, or other combinations of three final-state fermions, with appropriate substitutions. The only assumption required for our framework to hold is that the particle(s) mediating the MF or DF decay are massive enough to be integrated out, yielding a dimension-six four-fermion contact interaction.

We consider the following most-general four-fermion interaction Lagrangian,
\begin{equation}\label{eq:Lagrangian}
-{\cal L}_{\rm int} = \sum_{N,L,\alpha\beta} \left( G_{NL}^{\alpha\beta} \left[ \bar{\nu} \Gamma_N N\right] \left[ \bar{\ell}_\alpha \Gamma_L \ell_{\beta}\right] + \overline{G}_{NL}^{\alpha\beta} \left[ \bar{N} \Gamma_N \nu\right] \left[ \bar{\ell}_\alpha \Gamma_L \ell_{\beta}\right]\right) + h.c.,
\end{equation}
where the gamma matrices $\Gamma_N$ and $\Gamma_L$ are defined to include all possible interactions in this four-fermion structure and the indices $\alpha\beta=ee,\mu\mu,\tau\tau,e\mu,e\tau,\mu\tau$.\footnote{$G^{\alpha\beta}, \overline{G}^{\alpha\beta}, G^{\beta\alpha}, \overline{G}^{\beta\alpha}$ are coefficients to only two independent interactions in the MF case.} In practice, however, due to the available production mechanisms, we will mostly ignore $\tau$-lepton final-states.  In what follows we will often be restricting to a particular choice of lepton flavors and will suppress the $\alpha\beta$ indices for notational convenience.

We express the gamma matrices in terms of their Lorentz representation,
\begin{align}
\Gamma_{N},\ \Gamma_L &\in \left\lbrace \mathds{1},\ \gamma^5,\ \gamma^\mu,\ \gamma^\mu\gamma^5,\ \sigma^{\mu\nu}\equiv\frac{i}{2}\left[\gamma^\mu,\gamma^\nu\right]\right\rbrace,
\end{align}
and we will use the subscripts ``$SPVAT$'' to refer to scalar, pseudoscalar, vector, axial-vector, and tensor representations, respectively.  Lorentz invariance means that there are only 9 possible interaction structures and  we allow for interference among the different terms in our calculations.
The effective Lagrangian of Eq.~\eqref{eq:Lagrangian} is $U(1)_{EM}$ gauge invariant but not $SU(2)_L\times U(1)_Y$ invariant. It can, of course, be expressed as the low-energy limit of an $SU(2)_L\times U(1)_Y$ gauge-invariant effective Lagrangian.

Eq.~(\ref{eq:Lagrangian}) is valid if the HNL and the neutrino are both DFs or MFs. In the MF-case, however, the $\nu$ and $N$ are four-component Majorana fields. Under these conditions, the fermion bilinears $\bar{\nu} \Gamma_N N$ and $\bar{N} \Gamma_N \nu$ are related: $\bar{\nu} \Gamma_N N = \zeta_N \bar{N} \Gamma_N \nu$, where $\zeta_N=+1$ for $N=S,P,A$ and $\zeta_N=-1$ for $N=V,T$. Hence, if $N$ is a MF, the independent couplings\footnote{The relative minus sign comes from the fact that, when considering matrix elements of spinors instead of fields in a Lagrangian, the $\zeta_N$ flip sign. For example, $\bar{\nu} N = \bar{N} \nu$, but $\overline{u}_\nu u_N = -\overline{v}_N v_\nu$, which will enter our calculations of decay widths.} are $G_{NL}^{\alpha\beta}  - \zeta_N\overline{G}_{NL}^{\alpha\beta}$. For pragmatic reasons, we use the couplings defined in Eq.~(\ref{eq:Lagrangian}) to describe both MF and DF neutral leptons. 

While Eq.~(\ref{eq:Lagrangian}) is identical for DFs and MFs, the rest of the Lagrangian is not. Ignoring light neutrino masses, the bi-linear mass part of the Lagrangian is
\begin{equation}
{\cal L}_{\rm mass} = \frac{1}{2}m_N\bar{N}^cN,
\label{eq:MMass}
\end{equation}
for Majorana fermions and 
\begin{equation}
{\cal L}_{\rm mass} = m_N\bar{N}N,
\label{eq:DMass}
\end{equation}
for Dirac fermions. The number of degrees of freedom is different for MF (two) and DF (four) HNLs. The most general Lagrangian that contains four HNL degrees of freedom contains Eq.~(\ref{eq:DMass}) and $m'\bar{N}^c(a+b\gamma_5)N$ (these contain both ``left-left'' and ``right-right'' Majorana masses). For $m'\neq 0$, the Lagrangian describes, generically, two MF HNLs. The DF choice $m'=0$ is protected by a $U(1)$ lepton-number symmetry.  Here we only consider the one-Majorana-fermion case -- Eq.~(\ref{eq:MMass}) -- or the one-Dirac-fermion case -- Eq.~(\ref{eq:DMass}). 

While we keep the ordering of the fermion fields in Eq.~\eqref{eq:Lagrangian} in the ``neutral-current'' ordering, this parametrization is general. Since $\Gamma_N$ and $\Gamma_L$ span a complete basis, this allows for any type of ultraviolet completion with a charged and/or neutral force-carrier(s) mediating the $N$ decay.  Depending on the UV completion, not all the couplings presented in (\ref{eq:Lagrangian}) will be independent, and several can contribute to the same physical process.  
For instance, a DF which undergoes the decay $N\rightarrow \nu \ell^+_\alpha\ell_\alpha^-$ will receive contributions from both $G_{NL}^{\alpha\alpha}$ and $\left(\overline{G}_{NL}^{\alpha\alpha}\right)^*$. 
As we discuss below, instead of focusing on the terms in the Lagrangian, it is more convenient to structure the calculation in terms of the possible matrix elements associated to $N$ decay.

Throughout, we will concentrate on a few examples, including the special cases already highlighted: $\alpha=\beta$ and experimental setups that cannot distinguish $\ell^+$ from $\ell^-$. We will also discuss what happens when the $N$ decay is mediated by SM interactions. In this case, the effective Lagrangian for both Majorana and Dirac $N$ is
\begin{equation}\label{eq:BenchmarkLagrangian}
-{\cal L}_{\rm int} = 2\sqrt{2}G_F\left[ \bar{\nu} \gamma_{\mu}P_L N\right] \left[ \bar{\ell}_\alpha \gamma^{\mu}(g_L^{\alpha\beta}P_L+g_R^{\alpha\beta}P_R) \ell_{\beta}\right] + h.c.,
\end{equation}
where $P_{L,R}=(1\mp\gamma_5)/2$. $g_L$ and $g_R$ depend on the coupling between $N$ and the $W$ and $Z$ bosons and the light neutrinos. $g^{\alpha\beta}_R \propto \delta_{\alpha\beta}$, while the value of $g^{\alpha\beta}_L$, and whether it vanishes when $\alpha\neq\beta$, depends on the existence of a coupling between $N$, the $W$-boson, and $\ell_{\alpha}$.

If we further assume that all of these couplings are generated by $N$ mixing with one light neutrino, e.g. $\nu_\mu$ with a mixing angle $U_{\mu N}$, then the couplings may be determined: we give them in Table~\ref{table:gLgRDefs}.
\begin{table}[!h]
\begin{center}
\caption{Couplings $g_L$ and $g_R$ that enter the Lagrangian in Eq.~\eqref{eq:BenchmarkLagrangian} and Matrix Elements in Eq.~\eqref{eq:benchmarkM2} assuming that $N$ only mixes with $\nu_\mu$ with mixing angle $U_{\mu N}$. Here, $s_w^2 = 0.223$ is the (sine-squared of the) Weinberg angle.\label{table:gLgRDefs}}
\begin{tabular}{c|c||c|c}
Scenario & Example Process & $g_L$ & $g_R$ \\ \hline\hline
Distinct Final-State Charged Leptons  & $N \to \nu_e \mu^- e^+$ & $|U_{\mu N}|$ & $0$ \\ \hline
\multirow{2}{*}{Identical Final-State Charged Leptons} & $N \to \nu_\mu e^- e^+$ & $-\frac{1}{2}|U_{\mu N}|\left(1 - 2 s_w^2\right)$ & $|U_{\mu N}| s_w^2$ \\
 & $N \to \nu_\mu \mu^- \mu^+$ & $\frac{1}{2} |U_{\mu N}| \left(1 + 2s_w^2\right)$ & $|U_{\mu N}| s_w^2$
 \end{tabular}
\end{center}
\end{table}

\subsection{Matrix Elements}

We first consider the decay amplitude $\mathcal{M}_1$ of a spin-polarized DF $N$ into a light neutrino DF $\nu$ and two charged leptons $\ell_\alpha^-$ and $\ell_\beta^+$.
The matrix element can be written as
\begin{equation}\label{eq:MN}
\mathcal{M}_1= G_{NL} \left[ \overline{u}_\nu \Gamma_N P_S u_N \right] \left[ \overline{u}_\alpha \Gamma_L v_\beta\right],
\end{equation}
where $P_S \equiv \frac{1}{2}(1 + \gamma^5 \slashed{s})$ is a spin-projection operator\footnote{Here, $s^\mu$ is the spin of $N$, defined such that $s^2 = -1$ and $(p_N \cdot s) = 0$. Its spatial component points in the direction of the (assumed-to-be-polarized) $N$'s spin.}.
Above, and henceforth, we suppress the lepton flavor indices on $G_{NL}$.
In $\mathcal{M}_1$ 
there are \textit{nine} independent complex $G_{NL}$, \ie\ 18 free parameters, dictating this decay.

If $\mathcal{M}_1$ describes the matrix element for the decay $N \to \nu \ell^-_\alpha \ell^+_\beta$, we may write a related matrix element that describes the decay (if $N$ is a DF) $\overline{N} \to \overline{\nu} \ell^-_\alpha \ell^+_\beta$,
\begin{equation}\label{eq:MNBar}
\mathcal{M}_{2} = \overline{G}_{NL} \left[ \overline{v}_N P_S \Gamma_N v_\nu \right] \left[ \overline{u}_\alpha \Gamma_L v_\beta\right].
\end{equation}
The matrix element $\mathcal{M}_2$ is related to $\mathcal{M}_1$ by charge-conjugating the $N$/$\nu$ portion of the corresponding Feynman diagram, while the charged lepton piece remains untouched. 
$G_{NL}$ and $\overline{G}_{NL}$ can be completely unrelated: if $N$ is a DF with ``muon lepton number'', then the decays of $N$ will always produce a $\mu^-$ and the decays of $\overline{N}$ will always produce a $\mu^+$. If we are interested in final states with $\mu^+e^-$, then $\mathcal{M}_1$ will not contribute to any decay, while $\mathcal{M}_2$ will.\footnote{The ordering of spinors in the charged lepton leg in Eq.~(\ref{eq:MNBar}) is the same as in Eq.~(\ref{eq:MN}) -- the outgoing negatively-charged lepton is labelled ``$\alpha$'' and the positively-charged one is labelled ``$\beta$''. We will be interested in the decays of Dirac fermion $N$, $\overline{N}$, and Majorana fermion $N$ when comparing identical final states, such as $N \to \nu \mu^- e^+$ and $\overline{N} \to \overline{\nu} \mu^- e^+$, so the ordering of these labels is important.}

The Hermitian conjugate of the Lagrangian that leads to the decays to the $\ell_\alpha^-\ell_\beta^+$ final state will contribute to the charge-conjugated final state ($\ell_\alpha^+\ell_\beta^-$). Specifically, couplings proportional to $G_{NL}^*$ will contribute to $\overline{N}$ decay and couplings proportional to $\overline{G}_{NL}^*$ will contribute to $N$ decay. In the case where $\alpha\neq\beta$, but we have a charge-blind detector (one that can distinguish muons from electrons, but not $\mu^-$ from $\mu^+$ or $e^-$ from $e^+$), we must consider all of these contributions summed incoherently.

When writing Eqs.~(\ref{eq:MN}) and~(\ref{eq:MNBar}) in terms of four-spinors, we have adopted the canonical matrix element expression assuming that $N$ and $\nu$ are Dirac fermions. In the case that $N$ and $\nu$ are Majorana fermions, however, both $\mathcal{M}_1$ and $\mathcal{M}_2$ contribute to the decay $N \to \nu \ell^-_\alpha \ell^+_\beta$.  This can be seen at the level of the Lagrangian where $G_{NL}^{\alpha\beta}$ and $\overline{G}_{NL}^{\alpha\beta}$ both allow Majorana fermion $N$ to decay to the same final state.
Furthermore, the matrix elements for $\mathcal{M}_1$ and $\mathcal{M}_2$ are proportional up to an overall sign that depends on the gamma-matrix structure.  In the standard literature, e.g., Refs.~\cite{Formaggio:1998zn,Gorbunov:2007ak,Ballett:2019bgd}, Majorana fermion decay distributions and rates are calculated by taking the (non-interfering) matrix-elements-squared from Eqs.~(\ref{eq:MN}) and~(\ref{eq:MNBar}). This is valid in these works, when chiral projection operators $P_L$ and $P_R$ are acting on all spinors associated with $\nu$, causing any interference between these amplitudes to vanish (in the limit where the mass of $\nu$ is zero). When we allow for more generic $G_{NL}$ and $\overline{G}_{NL}$ however, interference can occur.

\textbf{Restrictions if $\boldsymbol{\alpha=\beta}$:} The Hermitian conjugate of the Lagrangian  that generates matrix elements $\mathcal{M}_1$ and $\mathcal{M}_2$ generates two new matrix elements, related to $\mathcal{M}_1$ and $\mathcal{M}_2$ by charge conjugation, which we refer to as $\mathcal{M}_1^c$ and $\mathcal{M}_2^c$. These are
\begin{eqnarray}
\mathcal{M}_1^c &= 
\eta_{N}\eta_{L} G_{NL}^* \left[ \overline{v}_N P_S \Gamma_N v_\nu\right] \left[ \overline{u}_\beta \Gamma_L v_\alpha\right], \label{eq:M1c}\\
\mathcal{M}_2^c &= 
\eta_{N}\eta_{L} \overline{G}_{NL}^* \left[\overline{u}_\nu \Gamma_N P_S u_N\right] \left[ \overline{u}_\beta \Gamma_L v_\alpha\right].\label{eq:M2c}
\end{eqnarray}
The prefactors $\eta_N$ and $\eta_L$ take into account the properties of the Lorentz structures $\Gamma_N$ and $\Gamma_L$ under conjugation, with $\eta_X = +1$ for $X = S, V, A, T$ and $-1$ for $X = P$.
If the final-state charged-leptons are identical, $\alpha=\beta$ and these new matrix elements contribute to the same process as $\mathcal{M}_{1,2}$.  If $N$ is a DF,  the decays of $N$ must be calculated using the (interfering) sum of $\mathcal{M}_1$ and $\mathcal{M}_2^c$: the decay rate of $N$ is proportional to $|\mathcal{M}_1 + \mathcal{M}_2^c|^2$ and the decay rate of $\overline{N}$ is $\propto |\mathcal{M}_2 + \mathcal{M}_1^c|^2$.

\textbf{Majorana Fermion $\boldsymbol{N}$ with $\boldsymbol{\alpha=\beta}$:} If the two charged fermions are of the same flavor then all four matrix elements contribute and the decay rate of $N$ is $\propto |\mathcal{M}_1 +\mathcal{M}_2 + \mathcal{M}_1^c+ \mathcal{M}_2^c|^2$.  
Furthermore, the different ordering of spinors in each matrix element leads to various cancellations when the sum is performed for a Majorana fermion $N$ decaying into a neutrino and identical final-state charged leptons. 
We denote the four possible combinations of couplings as
\begin{equation}
R_{NL}^\pm = \mathrm{Re}(G_{NL}\pm \overline{G}_{NL}),~~ I_{NL}^\pm = \mathrm{Im}(G_{NL}\pm \overline{G}_{NL})~.
\label{eq:ABReIm}
\end{equation}
The nine combinations that appear in the expression for the matrix element squared are $I_{SS}^-$, $R_{SP}^-$, $R_{PS}^-$, $I_{PP}^-$, $R_{VV}^+$, $R_{VA}^+$,  $I_{AV}^-$, $I_{AA}^-$, $R_{TT}^+ $.  The final expressions for the matrix element in this case are given in Table~\ref{table:MSqMajAB}, in Appendix~\ref{appendix:MSqs}.

\textbf{Restrictions if all new mediators are neutral:}
 A well motivated model is where the new-physics particles which mediate the $N$ decay are neutral\footnote{An example of this case is where a new $Z^\prime$ boson, heavier than $N$, is included to induce decays like $N \to \nu e^+ e^-$ via an off-shell $Z^\prime$.}. Since we focus on $m_N \le \mathcal{O}(\mathrm{GeV})$, which can be probed in fixed-target environments, strong constraints on new, charged mediators below the electroweak scale exist, and charged mediators with masses above the electroweak scale would likely provide small contributions to decays of this type relative to either those from the SM electroweak bosons or a new, light, neutral mediator.

Integrating out the neutral mediators present in a UV completion of this type induces relationships between the Lagrangian couplings in Eq.~(\ref{eq:Lagrangian}) and consequently between the parameters in matrix elements.  For example,
\begin{align}
|G_{NL}| = |\overline{G}_{NL}| \quad  & \mathrm{and} \quad \frac{G_{NL}}{\overline{G}_{NL}^*} = \eta_{N}\eta_{L}  \frac{G_{N'L}}{\overline{G}_{N'L}^*} ~,
\end{align}
where in the second relationship, $N'$ is the ``other'' $\Gamma$ matrix with which $\Gamma_{N}$ can interfere, e.g. if $N = V$, then $N^\prime = A$, etc. We provide further details about what occurs under this neutral mediator assumption in Appendix~\ref{appendix:NMO}.

\subsection{Lorentz invariants} \label{subsec:LIs}

When calculating the matrix-elements-squared, 
we find it useful to express our results as a linear combination of thirteen different\footnote{We discard terms proportional to $m_\nu$.} Lorentz-invariant quantities, which we label  as $K_j$ ($j = 1, 2, \ldots 13$). The subscript ``$m$'' refers to the negatively-charged lepton $\ell_\alpha^-$ and ``$p$'' to the positively-charged lepton $\ell_\beta^+$. Likewise, $m_m$ ($m_p$) is the mass of the negatively-(positively-)charged lepton. Six of these Lorentz-invariant quantities are $N$-spin independent,
\begin{align}
K_1 &= m_m m_p \left(p_\nu p_N\right),&  K_2 &= m_m m_N \left(p_p p_\nu\right),&  K_3 &= m_p m_N \left(p_m p_\nu\right), \label{eq:K1} \\
K_4 &= \left(p_p p_N\right) \left(p_m p_\nu\right),&  K_5 &= \left(p_m p_N\right) \left(p_p p_\nu\right),& K_6 &= \left(p_p p_m\right) \left(p_\nu p_N\right)~.
\end{align}
The remaining seven are $N$-spin dependent, 
\begin{align}
K_7 &= m_N \left(p_p p_m\right) \left(p_\nu s\right),& K_8 &= m_N m_m m_p \left(p_\nu s\right), \\
K_9 &= m_N \left(p_m p_\nu\right) \left(p_p s\right),& K_{10} &= m_N \left(p_p p_\nu \right) (p_m s),\\
K_{11} &= m_m \left[ \left(p_\nu p_N\right) \left(p_p s\right) - \left(p_p p_N\right) \left(p_\nu s \right)\right],& K_{12} &= m_p \left[ \left(p_\nu p_N\right) \left(p_m s\right) - \left(p_m p_N\right) \left(p_\nu s\right)\right],\\
K_{13}&= \varepsilon_{\rho\sigma\lambda\eta} p_m^\rho p_p^\sigma p_\nu^\lambda s^\eta.\label{eq:K13}
\end{align}
Note that except for $K_{13}$ all have mass-dimension four. This choice of Lorentz-invariants is not unique 
but it does have the benefit that the $K_i$ have simple properties under integration over subsets of the full phase space, as we discuss in Appendix~\ref{app:KinIntegration}.

Given this set of Lorentz-invariants, any matrix element squared describing $N$ decay can be expressed as
\begin{equation}\label{eq:MSqDecomposed}
\left\lvert \mathcal{M}(N \to \nu \ell_\alpha^- \ell_\beta^+)\right\rvert^2 = \sum_{j=1}^{13} C_j K_j ~.
\end{equation}

In Appendix~\ref{appendix:MSqs}, we provide the $C_j$ for Dirac fermion and Majorana fermion $N$ decays of the type $N \to \nu \ell^-_\alpha \ell^+_\beta$. We also analyze a few specific examples: when the physics mediating these decays is of the neutral-current variety and when the final-state charged leptons are identical ($\alpha = \beta$).  

In Eq.~\eqref{eq:BenchmarkLagrangian}, we introduced the Lagrangian of interest when the decays of $N$ are mediated by SM interactions. In this case,
\begin{eqnarray}
\mathcal{M}_1 &= 2\sqrt{2}G_F\left[ \overline{u}_\nu \gamma^\mu P_L P_S u_N \right] \left[ \overline{u}_\alpha \gamma_\mu \left(g_L P_L + g_R P_R \right) v_\beta\right], \label{eq:benchmarkM1}\\
\mathcal{M}_2 &= 2\sqrt{2}G_F\left[ \overline{v}_N P_S \gamma^\mu P_L v_\nu \right] \left[ \overline{u}_\alpha \gamma_\mu \left(g'_L P_L + g'_R P_R \right)v_\beta\right]~.
\label{eq:benchmarkM2}
\end{eqnarray}
The couplings $g'_{(L,R)}$ are in principle related to $g_{(L,R)}$ (see, for instance, Table~\ref{table:gLgRDefs}). If the final-state charged leptons are of the same flavor then $g_{(L,R)}=g'_{(L,R)}$. If the final-state charged leptons are distinct, then only a charged-current diagram with a $W$-boson contributes, and $g_{R} = g_R^\prime = 0$. The matrix elements in Eqs.~\eqref{eq:benchmarkM1} and \eqref{eq:benchmarkM2} can be mapped onto the language of Eqs.~\eqref{eq:MN} and~\eqref{eq:MNBar} using
\begin{align}
G_{VV} &= - G_{AV} = G_F \left(g_L + g_R\right)/\sqrt{2}, \label{eq:GVVBenchmark}\\
G_{VA} &= - G_{AA} = G_F\left(g_R - g_L\right)/\sqrt{2}, \\
\overline{G}_{VV} &= - \overline{G}_{AV} = G_F\left(g'_L + g'_R\right)/\sqrt{2}, \\
\overline{G}_{VA} &= - \overline{G}_{AA} = G_F\left(g'_R - g'_L\right)/\sqrt{2}.\label{eq:GVABarBenchmark}
\end{align}

For concreteness, we explore the scenario in which the decays of $N$ all arise due to $N$ mixing with $\nu_\mu$, via a ``mixing angle'' $U_{\mu N}$. Two specific considerations are worth exploring -- whether the final-state charged leptons are identical or not. 

In Table~\ref{tab:benchmarkMuE} we explore the case where they are not identical, with final-states being $\mu^- e^+$ or $\mu^+ e^-$. If $N$ only mixes with $\nu_\mu$ and lepton number is conserved, then a Dirac fermion $N$ can only decay into $\mu^- e^+$ and a Dirac fermion $\overline{N}$ into $\mu^+ e^-$. In these cases $g_L, g_L^\prime \to \left\lvert U_{\mu N}\right\rvert$ and $g_R, g_R^\prime \to 0$, as demonstrated in Table~\ref{table:gLgRDefs}. The only Lorentz invariants that appear in these cases are $K_4$, $K_5$, $K_9$, and $K_{10}$.

If the final-state charged leptons are identical, we find it simpler to provide results of $C_i$ in terms of $g_L$ and $g_R$ and present them in Table~\ref{tab:benchmarkAB}. The primed couplings are identical to the unprimed ones and are determined by the mixing $U_{\mu N}$ along with SM couplings  given in Table~\ref{table:gLgRDefs}. The difference between $g_L$ for $e^+ e^-$ and $\mu^+ \mu^-$ final states comes from the fact that, for the $\mu^+ \mu^-$ final state, the $W$-boson diagram interferes with the $Z$-boson one, whereas for $e^+ e^-$, only the $Z$ contributes. For Majorana fermion $N$ decays, there are relationships among the $C_i$, such as $C_4 = C_5$, $C_8 = 0$ and $C_9 = -C_{10}$.  These are directly related to the resulting lack of forward/backward asymmetry of Majorana fermion $N$ decays.

\begin{table}
\begin{center}
\caption{Lorentz-invariant contributions $C_i$ for decays into final states including $\mu^- e^+$ (left) and $\mu^+ e^-$ (right) under our benchmark model where decays are generated by mixing between $N$ and $\nu_\mu$ with mixing angle $U_{\mu N}$. All other $C_i$ are zero, and a common $G_F^2$ is factored out of each $C_i$.\label{tab:benchmarkMuE}}
\begin{tabular}{c | c| c|c||c|c|c} \hline \hline
& \multicolumn{3}{c||}{Final-state $\mu^- e^+$} & \multicolumn{3}{c}{Final-state $\mu^+ e^-$} \\ \hline
$C_i$ & Dirac $N$ & Dirac $\overline{N}$ & Majorana $N$ & Dirac $N$ & Dirac $\overline{N}$ & Majorana $N$ \\ \hline
$C_4$ & $64 \absq{U_{\mu N}}$ & 0 & $64\absq{U_{\mu N}}$ & 0 & 0 & 0 \\
$C_5$ & $0$ & 0 & 0 & 0 & $64\absq{U_{\mu N}}$ & $64 \absq{U_{\mu N}}$ \\ \hline
$C_9$ & $-64 \absq{U_{\mu N}}$ & 0 & $-64 \absq{U_{\mu N}}$ & 0 & 0 & 0 \\ 
$C_{10}$ & $0$ & 0 & 0 & 0 & $64\absq{U_{\mu N}}$ & $64\absq{U_{\mu N}}$ \\ \hline\hline
\end{tabular}
\end{center}
\end{table}

\begin{table}[!t]
\begin{center}
\caption{$C_i$ for simple benchmark model of $N$ decay into identical final-state charged-lepton states $N \to \nu \ell_{\alpha}^+ \ell_{\alpha}^-$, through $N-\nu_\mu$ mixing (see Eqs.~(\ref{eq:benchmarkM1}) and (\ref{eq:benchmarkM2})). All other $C_i$ are zero, and a common $G_F^2$ is factored out of each $C_i$. Here, $g_L = \left\lvert U_{\mu N}\right\rvert\delta_{\mu\alpha} - \frac{1}{2}\left\lvert U_{\mu N}\right\rvert(1 - 2s_w^2)$ and $g_R = \left\lvert U_{\mu N}\right\rvert s_w^2$. \label{tab:benchmarkAB}}
\begin{tabular}{c|ccc}\hline\hline
$C_i$ & Dirac $N$ & Dirac $\overline{N}$ & Majorana $N$ \\ \hline
$C_1$ & $64 g_L g_R$ & $64 g_L g_R$ & $128 g_L g_R$ \\
$C_4$ & $64 g_L^2$ & $64 g_R^2$ & $64\left(g_L^2 + g_R^2\right)$ \\
$C_5$ & $64 g_R^2$ & $64 g_L^2$ & $64\left(g_L^2 + g_R^2\right)$ \\ \hline
$C_8$ & $-64 g_L g_R$ & $64 g_L g_R$ & $0$ \\
$C_9$ & $-64g_L^2$ & $64g_R^2$ & $64\left(g_R^2 - g_L^2\right)$ \\
$C_{10}$ & $-64g_R^2$ & $64g_L^2$ & $-64\left(g_R^2 - g_L^2\right)$ \\ \hline\hline
\end{tabular}
\end{center}
\end{table}


\section{Anisotropy of Dirac/Majorana Fermion Decays}
\label{sec:Anisotropy}
\setcounter{footnote}{0}

We now wish to make connection with the kinematic observables that have the potential to distinguish between the MF and DF hypotheses. We restrict our discussion to the $N$ rest-frame; recasting the discussion to the laboratory frame  is possible but often very cumbersome, especially since we are interested in three-body decays. If $N$ production is through meson decay at rest (e.g. $\pi^+\to\mu^+ N$), it is simple to reconstruct the $N$ rest frame, and the results discussed here are readily applicable.

We will discuss the hypotheses that both the HNL and the light neutrino are either MF or DF, and assume the HNL is 100\% polarized unless otherwise noted.

\begin{figure}[htbp]
\begin{center}
\includegraphics[width=0.50\linewidth]{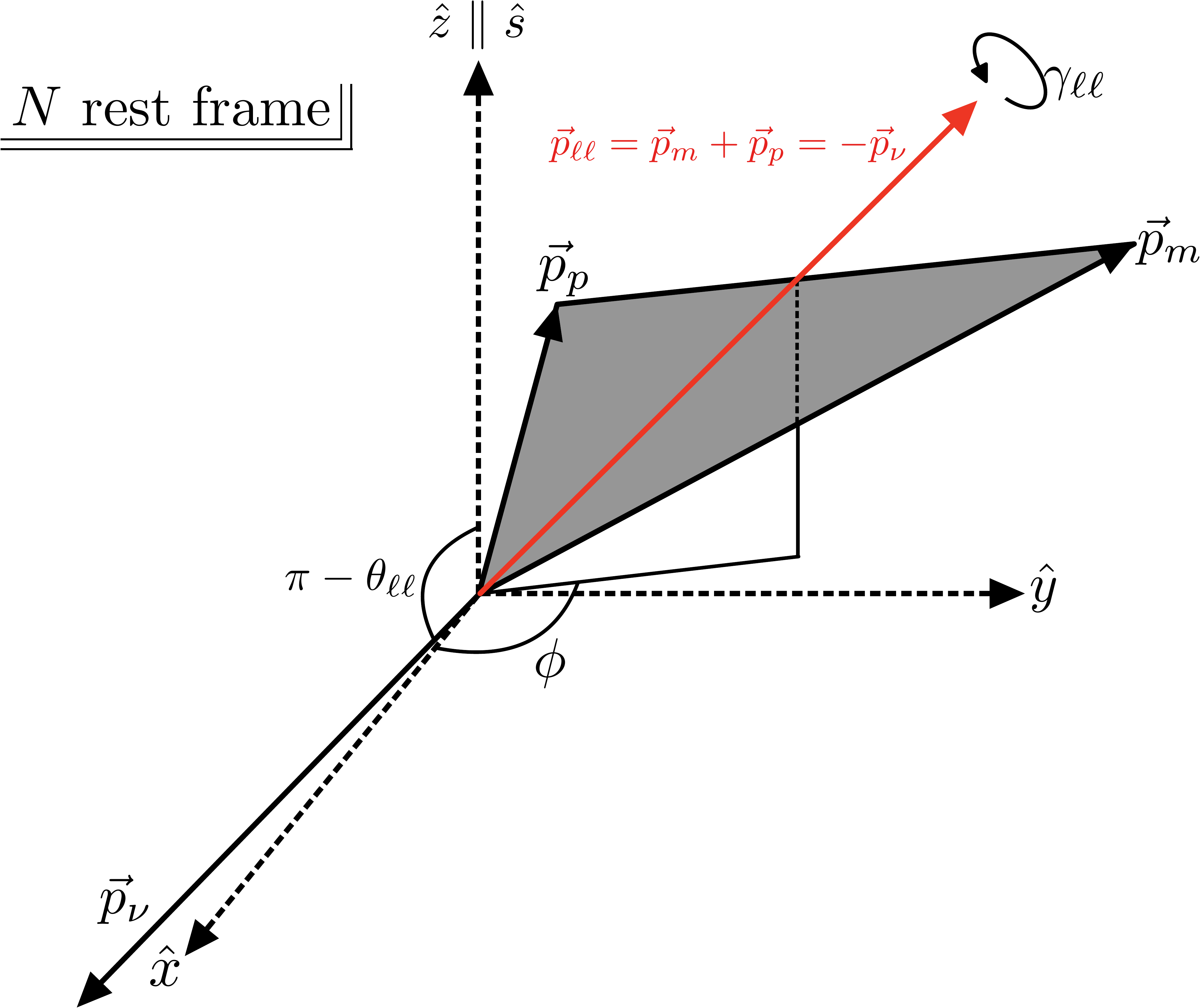}
\caption{Kinematics of the decay $N \to \nu (p_\nu) + \ell_\alpha^- (p_m) + \ell_\beta^+ (p_p)$ in the rest-frame of the decaying parent particle $N$. Black, solid arrows represent the three-momenta of the three final-state particles, and the red arrow is the sum of the two charged-lepton three-momenta, $\vec{p}_{\ell\ell}= \vec{p}_m + \vec{p}_p$. The three angles involved in the kinematics are as follows: 1) $\theta_{\ell\ell}$, the angle between the spin direction of $N$ (which defines the $\hat{z}$ direction) and $\vec{p}_{\ell\ell}$ -- in the diagram, we label $\pi - \theta_{\ell\ell}$ for convenience. 2) $\gamma_{\ell\ell}$, the angle that defines the orientation of the plane defined by $\vec{p}_{p}$ and $\vec{p}_{m}$ relative to $\vec{p}_{\ell\ell}$. 3) $\phi$, the azimuthal angle of $\vec{p}_{\ell\ell}$.\label{fig:Kinematics}}
\end{center}
\end{figure}
In the $N$ rest-frame, the momentum vectors of the three decay products are coplanar, and we parametrize their energies using $m_{\ell\ell}^2$, the invariant mass squared of the charged lepton system, and $m_{\nu m}^2$, the invariant mass squared of the neutrino and the negatively charged lepton $\ell_\alpha^-$.  
The orientation of the plane relative to the $N$ spin direction $\hat{s}$, which we take to be aligned with the $z$-axis, is defined by three angles: $\cos\theta_{\ell\ell}$, the (cosine of the) angle between the $N$ spin direction $\hat{s}$ and the direction of the outgoing charged-lepton pair, $\vec{p}_{\ell\ell} \equiv \vec{p}_\alpha + \vec{p}_\beta$, $\gamma_{\ell\ell}$, a rotation angle about $\vec{p}_{\ell\ell}$, defined such that $\gamma_{\ell\ell}=0$ corresponds to the vector normal to the decay plane being perpendicular to the $z$-axis, and an
azimuthal angle $\phi$, which corresponds to rotations of the entire system about the spin direction $\hat{s}$.

Using this choice of phase-space variables, the fully differential partial width for the $N$ decay is
\begin{equation}\label{eq:dGamma}
\frac{d\Gamma(N \to \nu \ell_\alpha^- \ell_\beta^+)}{d m_{\ell\ell}^2 d \cos\theta_{\ell\ell} d m_{\nu m}^2 d \gamma_{\ell\ell} d \phi} = \frac{1}{\left(2\pi\right)^5} \frac{1}{64m_N^3} \left\lvert \mathcal{M}\right\rvert^2  = \frac{1}{\left(2\pi\right)^5}  \frac{1}{64m_N^3} \sum_{j=1}^{13} C_j K_j.
\end{equation}
The expressions for $K_i$ in terms of the lab-frame phase space variables are given in Appendix~\ref{app:KinIntegration}.

In order to determine the ``geometric'' properties of the decay, including the forward/backward asymmetry and  the dependency of the differential width on $\gamma_{\ell\ell}$, we integrate over the non-angular variables $m_{\ell\ell}^2$ and $m_{\nu m}^2$ (a reminder that $\phi$ can be trivially integrated for all cases of interest). This procedure is detailed in Appendix~\ref{app:Integration}. The forward-backward asymmetry $A_{\rm FB}$ is defined in Eq.~\eqref{eq:FBasymmetry} and can be computed exactly, further integrating the differential width over $\gamma_{\ell\ell}$, for all of the decay models defined earlier. In Section~\ref{subsec:AllowedAnisotropyMajorana} and in Appendix~\ref{app:Integration}, we discuss the forward/backward asymmetry for certain test cases and explicitly confirm the results discussed in Section~\ref{sec:CPTArguments}. Namely, we verify that, if $N$ is a MF,
\begin{itemize}
\item $A_{\rm FB}=0$ if the final-state charged leptons are identical, $\alpha = \beta$.
\item $A_{\rm FB}=0$ if the experiment detecting the final-state particles is charge-blind, and cannot distinguish between the final states $\ell^-_\alpha \ell^+_\beta$ and $\ell^+_\alpha \ell^-_\beta$.
\end{itemize}
Previous results in the literature, such as Ref.~\cite{Formaggio:1998zn,Ballett:2019bgd}, concluded that Majorana fermion $N$ decays can have anisotropy (even in some of the cases listed above) because the decay distributions were analyzed in terms of the outgoing direction of a single charged lepton, as opposed to the charged lepton pair.

\subsection{Allowed Asymmetry for Dirac Fermion Decays}
\label{subsec:AllowedAnisotropyDirac}

While there are general circumstances where the decay of a MF $N$ is guaranteed to be forward-backward symmetric, the decay of a DF $N$ can be highly anisotropic. We discuss this in more detail in this subsection, concentrating on $A_{\rm FB}$.

$A_{\rm FB}\neq 0$ requires the presence of at least two types of couplings. If \emph{only} two types of couplings are nonzero, anisotropy can occur for scalar/pseudoscalar only or vector/axial-vector only. All other pairwise combinations result in zero $A_{\rm FB}$. These two combinations of operators would arise if $N$ was coupled to the SM through exchange of either a spin-0 boson or a spin-1 boson with generic couplings to the SM fermions.

\paragraph{Scalar/Pseudoscalar Interactions Only:} The forward-backward asymmetry with only scalar/pseudoscalar interactions is
\begin{equation}\label{eq:AlphaSPOnlyG}
\afb^{(SP)} = \frac{\mathrm{Re}\left( G_{PP} G_{SP}^* +G_{SS} G_{PS}^* \right)T_0 - \left[\delta^2 \mathrm{Re}\left(G_{PP} G_{SP}^*\right) + \sigma^2 \mathrm{Re}\left(G_{SS} G_{PS}^*\right) \right] T_1}{\left[|G_{SS}|^2 + |G_{SP}|^2 + |G_{PS}|^2 + |G_{PP}|^2\right] T_0 - \left[ \delta^2 \left(|G_{PP}|^2 + |G_{SP}|^2\right) +  \sigma^2 \left(|G_{SS}|^2 + |G_{PS}|^2\right)\right]T_1}~.
\end{equation}
Here $T_{0,1}$, computed in Appendix~\ref{eq:TIntegrals}, are dimensionless functions of the particle masses, while $\sigma,\delta$ are ratios of masses:
\begin{equation}
\sigma  \equiv \frac{m_m + m_p}{m_N},\quad \quad  \delta  \equiv \frac{m_m - m_p}{m_N}~.
\end{equation}
 
If the four-fermion couplings are generated by the exchange of neutral spin-0 bosons then they are related: $G_{SS} G_{PP} = G_{SP} G_{PS}$, and 
\begin{equation}\label{eq:AlphaSPOnlyNL}
\afb^{(\mathrm{SP})} = \frac{\mathrm{Re}\left(G_{PP} G_{SP}^*\right)}{\left\lvert G_{PP}\right\rvert^2 + \left\lvert G_{SP}\right\rvert^2}~.
\end{equation}

This fraction is extremized when $\left\lvert G_{PP}\right\rvert = \left\lvert G_{SP} \right\rvert$ and $\mathrm{Re}(G_{PP} G_{SP}^*) = \pm \left\lvert G_{PP}\right\rvert \left\lvert G_{SP}\right\rvert$ resulting in $A_{\rm FB}=\pm 1/2$.

\paragraph{Vector/Axial-vector Interactions Only:} The well-studied case in which $N$ interacts with the Standard Model only via mixing with light neutrinos falls into this class, and we discuss it in turn. In the case where only these couplings are nonzero,
\begin{equation}\label{eq:AnisotropyVA}
\afb^{(VA)} = \frac{ \mathrm{Re}( G_{VV} G_{AV}^* + G_{AA} G_{VA}^*) f_+ + \mathrm{Re}(G_{VV} G_{AV}^* - G_{AA} G_{VA} ^*) f_0}
{\left[|G_{VV}|^2 + |G_{AA}|^2 + |G_{VA}|^2 + |G_{AV}|^2\right] f_- +\left[|G_{VV}|^2 - |G_{AA}|^2 - |G_{VA}|^2 + |G_{AV}|^2\right] f_0}~,
\end{equation}
with 
\begin{align}
f_{\pm} & =2\left[ 4T_0 - \left( \pm2 + \sigma^2 + \delta^2 \right) T_1 - \left(\pm\sigma^2 \pm \delta^2 + 2 \sigma^2 \delta^2 \right)T_2 \pm 4\sigma^2\delta^2 T_3\right]~,\label{eq:fpm}\\
f_0 & = 6\left(\sigma^2-\delta^2\right) T_1~.\label{eq:f0}
\end{align}

In the limit $\sigma \to 1$, when the decay $N \to \nu \ell_\alpha^- \ell_\beta^+$ is barely kinematically accessible, this anisotropy parameter becomes
%
\begin{equation}
\afb^{(VA)} \to \frac{\mathrm{Re}\left(G_{VV} G_{AV}^* - G_{AA} G_{VA}^* \right)}{3 \absq{G_{VV}} + \absq{G_{AA}} + \absq{G_{VA}} + 3\absq{G_{AV}}}~,
\end{equation}
which can be as large in magnitude as $1/2$, corresponding to maximal anisotropy.

Instead, in the limit  where the final state masses can be ignored relative to the parent mass (if $\sigma,\delta\to 0$), 
\begin{equation}
\afb^{(VA)} \to -\frac{\mathrm{Re}\left(G_{AA} G_{VA} ^* + G_{VV} G_{AV}^*\right)}{3\left(\absq{G_{VV}} + \absq{G_{AA}} + \absq{G_{VA}} + \absq{G_{AV}}\right)},
\end{equation}
which can be as large in magnitude as $1/6$.

\begin{figure}
\begin{center}
\includegraphics[width=0.495\linewidth]{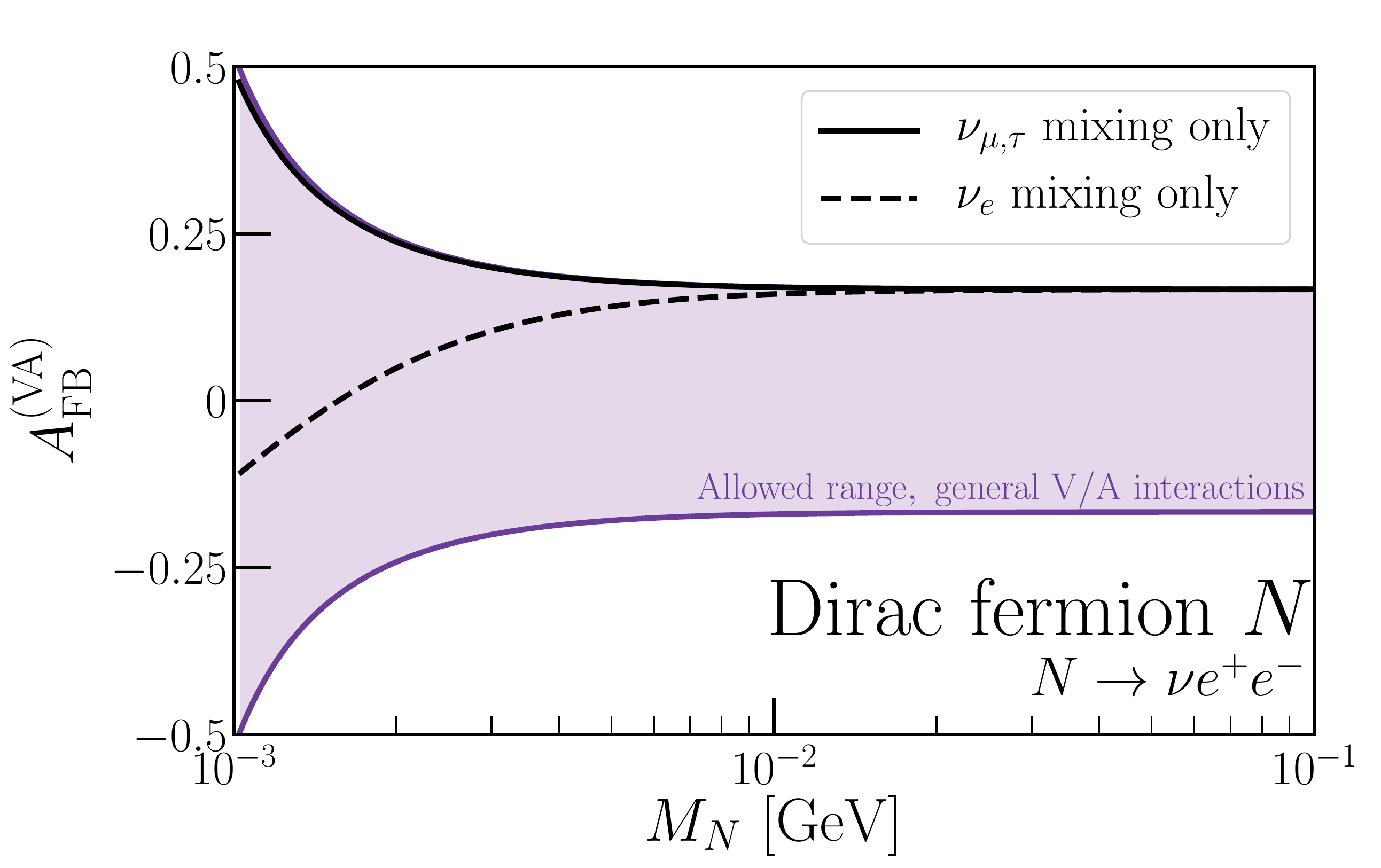}
\includegraphics[width=0.495\linewidth]{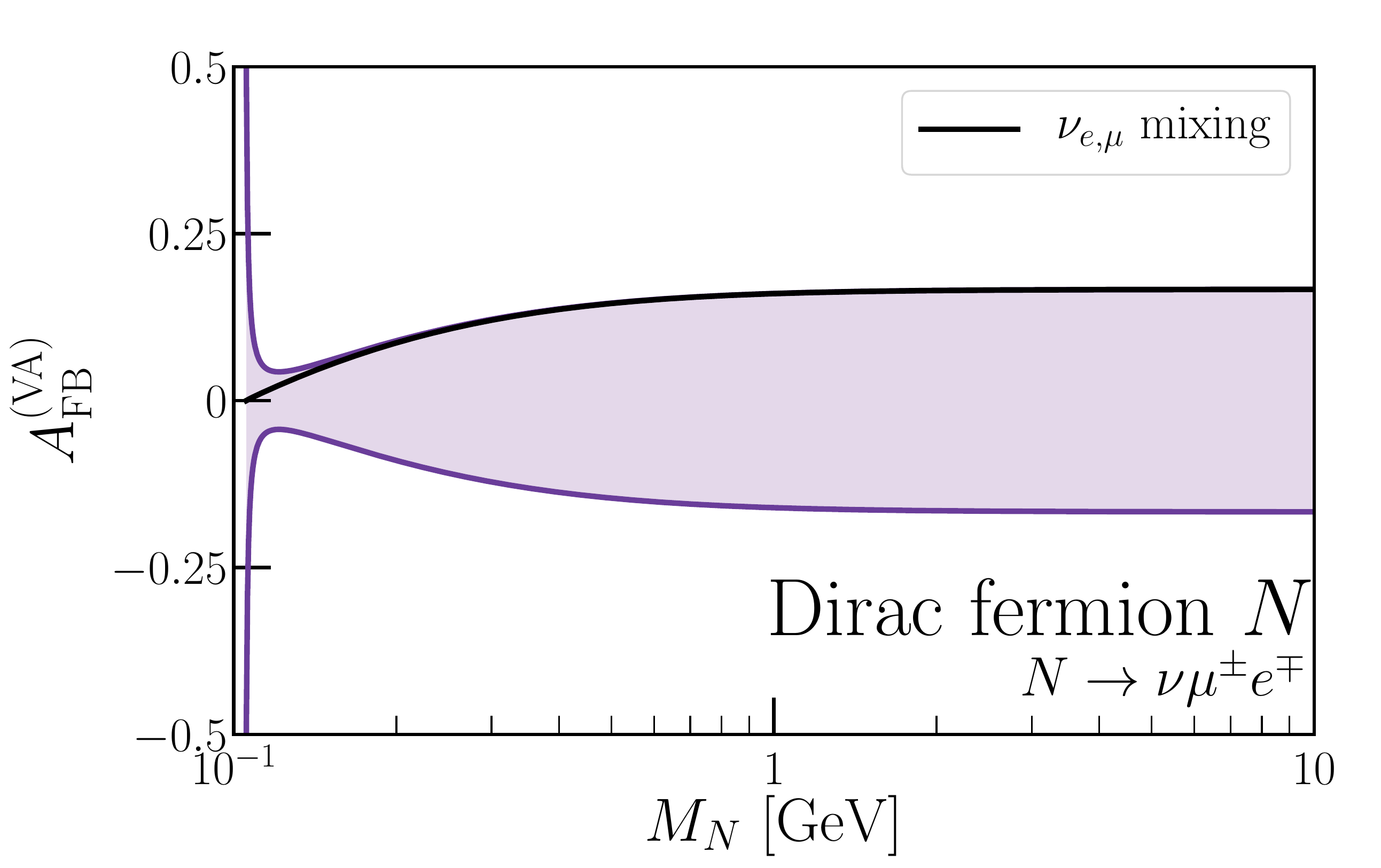}
\caption{Allowed range of the forward-backward asymmetry $A_{\rm FB}$ (shaded band) as a function of the DF HNL mass $m_N$ in the decay $N\to \nu e^+ e^-$ (left) or $N\to\nu e^\pm \mu^\mp$ (right), for general vector and axial-vector couplings for the neutrinos and charged leptons.  The case of decay through $N$ mixing with a single flavor of SM neutrino is depicted by the black solid ($U_{\mu N}$ or $U_{\tau N}$ nonzero) and dashed ($U_{eN}$ nonzero) lines. In the right-hand panel, the different cases overlap.
\label{fig:AlphaVA}}
\end{center}
\end{figure}

The allowed values of $A_{\rm FB}$ as a function of the mass of $N$ are depicted in Fig.~\ref{fig:AlphaVA}. The left panel is for the decay $N \to \nu e^- e^+$ and the right panel is for $N \to \nu \mu^- e^+$. In both cases, at threshold (low $m_N$), $A_{\rm FB}$ can take on values between $\pm 1/2$, while at large $m_N$ it is restricted to lie between $\pm 1/6$, as described in the limiting cases above.

Each panel of Fig.~\ref{fig:AlphaVA} also depicts $A_{\rm FB}$ for the benchmark model of decay through $\nu-N$ mixing, with the matrix element given in Eq.~\eqref{eq:benchmarkM1} and assuming that the $N$ mixes with only one flavor of SM neutrino, i.e., only one among the three $U_{\kappa N}\neq0$, $\kappa=e,\mu,\tau$.  For the $e^+ e^-$ final state (left panel), mixing with $\nu_e$ results in contributions from both $W$ and $Z$ exchange while mixing with $\nu_\mu$ or $\nu_\tau$ receives a contribution only from $Z$ exchange.  
In the case of the $e^\pm\mu^\mp$ final state (right panel) there is only the contribution from $W$ exchange.

\subsection{Allowed Asymmetry for Majorana Fermion Decays}
\label{subsec:AllowedAnisotropyMajorana}
In this subsection, we revisit the results of Section~\ref{sec:CPTArguments}, assisted by results in Appendices~\ref{appendix:MSqs} and~\ref{app:Integration}, to determine how large the forward-backward asymmetry can be for Majorana fermions in other specific cases.

In Appendix~\ref{app:Integration}, in determining how different Lorentz-invariant contributions $C_i$ lead to a forward/backward asymmetry, we introduced $X_{\rm FB}$, see Eqs.~(\ref{eq:XFBDef},\ref{eq:DoubleDiffAngleSimp}). $A_{\rm FB}$ is directly proportional to $X_{\rm FB}$, which can be written as 
\begin{equation}
\sum_{m = 0}^{3} \sum_{i=7}^{12}  C_i D^{i}_m T_m.
\end{equation}
The $D_m^{i}$ coefficients, $m=0,\ldots 3$, $i=7,\ldots 12$, are listed in Table~\ref{Tab:LIContributionsSD} and $\sum_{i=7}^{12} C_i D^{i}_m$ can be expressed as
\begin{align}
\sum_{i=7}^{12} C_i D^{i}_m &= \left( \begin{array}{c} 2 C_7 + \frac{2}{3} \left(C_9 + C_{10}\right) \\ \sigma^2 \left(C_8 - C_7\right) - \frac{1}{6}\left(2+\sigma^2\right)\left(C_9 + C_{10}\right) - \sigma\left(C_{11} + C_{12}\right) \\ -\frac{\sigma^2}{6}\left(C_{9} + C_{10}\right) \\ 0\end{array}\right) \nonumber \\
& + \delta \left( \begin{array}{c} 0 \\ -\delta\left(C_{7} + C_{8}\right) - \frac{\delta}{6}\left(C_{9} + C_{10}\right) + \left(C_{12} - C_{11}\right) \\ -\frac{\delta}{6}\left(1 + 2\sigma^2\right)\left(C_{9} + C_{10}\right) + \sigma\left(C_{11} \left(\sigma + \delta\right) - C_{12} \left(\sigma-\delta\right)\right) \\ \frac{2\sigma^2\delta}{3}\left(C_{9} + C_{10}\right) \end{array}\right).\label{eq:CDs}
\end{align}
Because this object is contracted with $T_m$ to determine $A_{\rm FB}$, then the forward-backward asymmetry is zero if each element of $\sum_{i=7}^{12} C_i D^{i}_m$ is zero. Cancellation could occur in the contraction between this object and $T_m$, however, if each element is zero then this will guarantee $A_{\rm FB} = 0$ for all possible combinations of charged lepton and $N$ masses. First, let us focus on the case where the final-state charged leptons are identical so $\delta = 0$ and the second line in Eq.(\ref{eq:CDs}) vanishes. By inspection, we see that  $\sum_{i=7}^{12} C_i D^{i}_m$ will vanish, for all $\sigma$, if the following are all true: $C_9 + C_{10} = 0$, $C_{11} + C_{12} = 0$, $C_{7} = C_{8} = 0$. This set of relations is realized -- see  Table~\ref{table:MSqMajAB} -- when $N$ is a Majorana fermion decaying into identical final-state charged leptons. 

Another case of interest is when only neutral mediators contribute to the decay of a Majorana fermion $N$, either into identical or distinct final-state charged leptons. In Table~\ref{table:MSqMajNMO}, we provide these results\footnote{The identical final-state charged lepton case, $\delta = 0$, is given in Table~\ref{table:MSqMajABNMO} and follows the pattern above, generating zero forward-backward asymmetry.}. In this case, we have $C_{7} = C_{8} = 0$ and $C_{9} + C_{10} = 0$. However, in the case of distinct charged-leptons in the final state $\delta \neq 0$ and we must keep both terms of Eq.~\eqref{eq:CDs}. In this case, the forward-backward asymmetry will be proportional to
\begin{align}
X_{\rm FB}^{\rm NMO} \to C_{11} \left(\sigma + \delta\right)\left( \sigma\delta T_2 - T_1\right) + C_{12} \left(\delta - \sigma\right)\left( \sigma\delta T_2 + T_1\right).\label{eq:XFBNMO}
\end{align}
Several features here are of note. First, only $C_{11}$ and $C_{12}$ contribute to the forward backward asymmetry when $N$ is a Majorana fermion decaying via neutral mediators only. These two coefficients require interference between $G_{NL}$ of the vector/axial-vector type with those of the scalar-pseudoscalar or tensor types. In other words, if we only have neutral mediators, multiple Lorentz representations must appear in the decay matrix element or mediators of different spin (spin-1 and spin-0 or spin-2) must contribute. Additionally, in Eq.~\eqref{eq:XFBNMO}, the factors multiplying $C_{11}$ and $C_{12}$ are proportional to $\sigma + \delta = 2m_{\ell_\alpha}/m_N$ and $\delta - \sigma = -2m_{\ell_\beta}/m_N$ so as $m_N$ grows relative to the masses of the charged leptons into which it is decaying, the forward backward asymmetry shrinks.

\begin{figure}[!htbp]
\begin{center}
\includegraphics[width=0.6\linewidth]{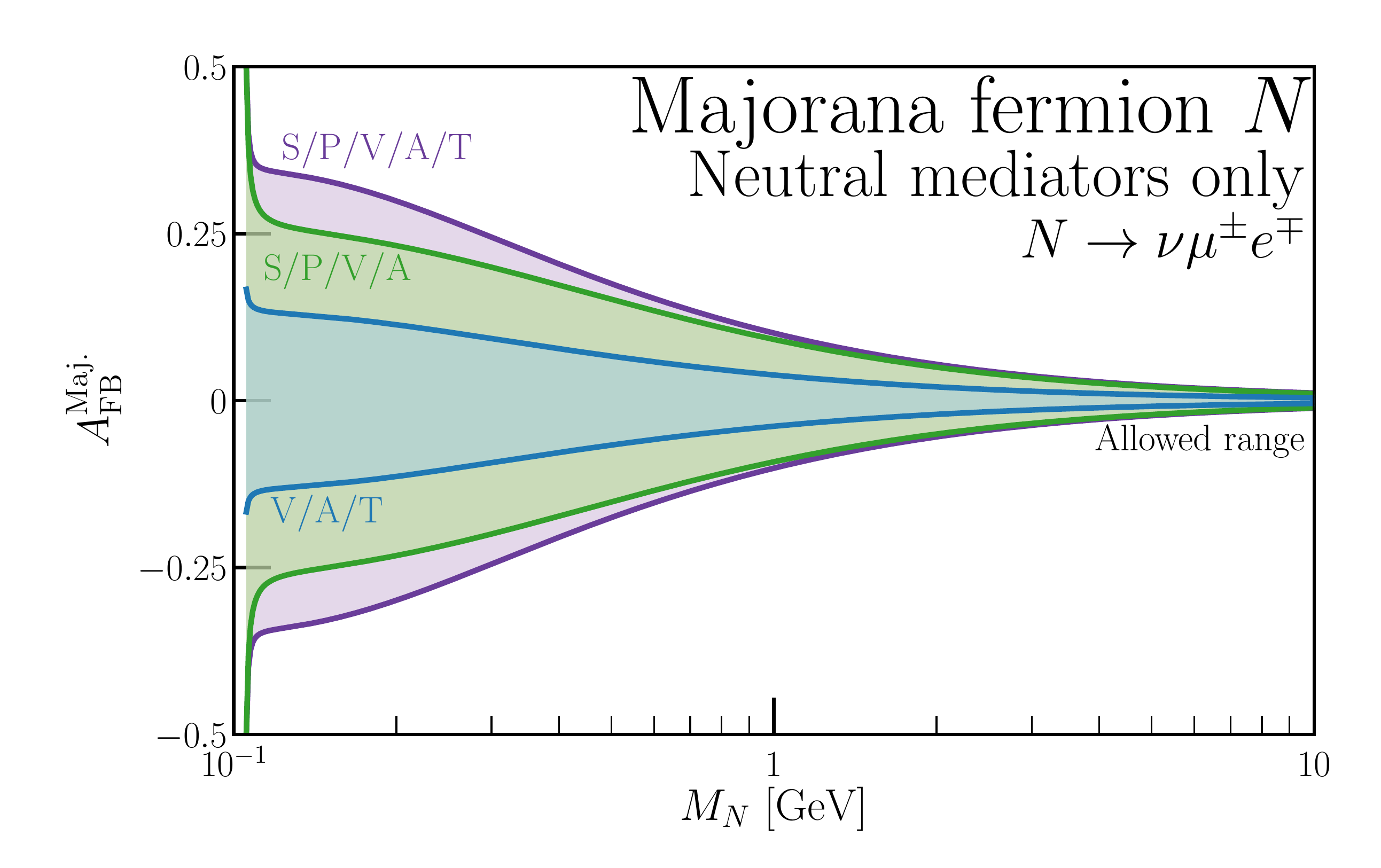}
\caption{Allowed range of decay anisotropy for the decay $N \to \nu \mu^- e^+$ or $\nu \mu^+ e^-$ when $N$ is a Majorana fermion. Here, we consider decays induced by neutral mediators only, and allow for nonzero couplings of the scalar, pseudoscalar, vector, and axial-vector variety (in green) or the vector, axial-vector, and tensor variety (in blue). The purple region indicates the allowed range when all types of couplings are allowed to be nonzero.
\label{fig:AlphaMaj}}
\end{center}
\end{figure}

In order to determine how large the forward-backward asymmetry can be, we explore the case where a Majorana fermion $N$ decays into $\nu \mu^- e^+$ or $\nu \mu^+ e^-$ via neutral mediators. 
We consider several combinations of allowed couplings, as labelled in Fig.~\ref{fig:AlphaMaj}. The largest asymmetry is attained when all couplings (SPVAT) are allowed to be nonzero, and for $m_N$ just larger than $m_\mu + m_e$.

\section{Dirac Fermion Mimicking Majorana Fermion Decay Distributions}
\label{sec:DiracFakingMajorana}

As we have seen from the previous sections, with more details given in Appendix~\ref{appendix:MSqs}, generically DF have significantly larger forward-backward anisotropies than MF. Additionally, this anisotropy must be zero for MF for a number of well-motivated model-dependent scenarios, such as HNLs that decay via a single neutral mediator only.  Despite these generic differences, it is still possible, in certain restricted cases, that both DF and MF hypotheses are capable of explaining observations.  We now address the circumstances under which the DF hypothesis can faithfully fake the MF one.

Instead of focusing only on the produced forward-backward asymmetry, we consider the full matrix-element-squared and therefore the fully-differential partial widths of the decays. If $N$ is a MF, can the parameters $G_{NL}$ associated with a DF $N$ conspire perversely such that this decay is perfectly mimicked in all ways (\ie\  the matrix-elements-squared become identical)? For any single decay channel this is always possible, since if $N$ is a MF there are additional restrictions on the form of the Lagrangian and the MF decay can be derived from the DF decay with specific substitutions, see Eqs.~\eqref{eq:appDirtoMajstart}-\eqref{eq:appDirtoMajend}. Under these substitutions, the coefficients of the different Lorentz-Invariant contributions assuming $N$ is a DF, given in Table~\ref{table:MSqDirac}, align with those for a MF, given in Table~\ref{table:MSqMajGen}.  Thus, if one can only observe a single decay channel of $N$ (\eg\ $N \to \nu \ell_\alpha^- \ell_\beta^+$) and there is no information on, for example, related final states (\eg\ $\nu\ell_\beta^- \ell_\alpha^+ $) and the data are consistent with the MF hypothesis, it is always possible to find a DF scenario which is also consistent: a DF can always fake a MF.  The converse is not true, a generic DF cannot always be faked by a MF. For example, there are many cases in which all couplings for a MF will lead to no asymmetry and a typical set of DF couplings leads to nonzero asymmetry.

If multiple decay channels are experimentally accessible, \eg\ $N \to \nu \mu^\pm e^\mp$, it becomes harder for the DF hypothesis to fake the MF one.  Regardless of couplings and mixing, the ratio of rates for a MF to decay into these two final states must be one (within statistical fluctuations).  For a DF, this ratio depends on both the relative production rate of $N$ and $\overline{N}$ (for instance, from positively-charged and negatively-charged meson decays into the $N$ or $\overline{N}$) as well as the parameters governing the decay of $N$ itself.  
Data that are consistent with the MF hypothesis may only be faked by a DF if the production and decay processes conspire.  Something as straightforward as altering production modes, by altering focussing or changing beam energy, should be able to break any degeneracy and distinguish these hypotheses. While it is possible for these effects to conspire in the DF scenario in a way that the relative rate of the two final-states is unity, in general considering both of these channels in tandem \textit{should} provide stronger evidence that $N$ is a Majorana fermion.

The situation is also different if one only sticks to concrete models for the physics behind HNL production and decay.
As a specific example, consider the case where the only interactions of $N$ are through mixing with the light neutrinos $\nu_\kappa$ with mixing angles $U_{\kappa N}$ so that all of $N$ decays are mediated by the SM weak interactions. For the same lepton-flavor final state, the forward-backward asymmetry is zero if $N$ is a MF.  Generically, a DF will have nonzero forward backward asymmetry and thus the two hypotheses can be distinguished from $A_{\rm FB}$ alone. In general, when all $U_{\kappa N}$ are relevant,  the matrix-element-squared for the decay of interest is 
\begin{equation}
\left\lvert \mathcal{M}(N \to \nu e^+ e^-)\right\rvert^2 = \sum_{\kappa = e,\mu,\tau} \left\lvert \mathcal{M}(N \to \nu_\kappa e^+ e^-)\right\rvert^2,
\end{equation}
where each of the three processes on the right-hand side will depend on an individual mixing angle $\left \lvert U_{\kappa N}\right\rvert^2$. As these mixings change, the allowed forward-backward asymmetry of the $e^+ e^-$ final state can vary between the values predicted by $\nu_e$ mixing alone and $\nu_{\mu,\tau}$ mixing \ie\ $A_{\rm FB}$ can take any value between the solid and dashed black lines of Fig.~\ref{fig:AlphaVA}(left).  For $m_N \lesssim 1.5$ MeV, this range includes zero and thus if the measurement in the decay $N \to \nu e^+ e^-$ is consistent with $A_{\rm FB}=0$ it is not possible to determine if $N$ is a Majorana fermion or Dirac fermion.  Furthermore, the DF hypothesis will require specific arrangements of $\left\lvert U_{\kappa N}\right\rvert^2$.  Conversely, if $m_N$ is sufficiently large (and we restrict ourselves to this SM-like scenario) and $A_{\rm FB}$ is measured to be zero, the evidence is in favor of $N$ being a Majorana fermion. 

We have implicitly assumed that (a) the $N$ are produced via some process from which they emerge 100\% polarized and (b) we are operating in a setting in which (if $N$ is a DF) we only have $N$ production and not $\overline{N}$. If either of these assumptions is violated, then the distinction between DF and MF becomes more difficult. For instance, if $N$ is a DF and $N$ and $\overline{N}$ are produced in equal abundance, their contributions cancel any forward/backward asymmetry, a signature indicative of a MF decay. In principle, as long as the net polarization of the $N$ is nonzero and the Dirac $N$/$\overline{N}$ production rates are not equal, the same mechanisms for separating the MF/DF hypotheses we discussed throughout this work are still accessible. However, the reduced separation between the two hypotheses will mean that larger statistical samples of $N$ decays are required to perform this distinction at a meaningful level. We will analyze this situation in detail in Ref.~\cite{deGouvea:2021TBA}.

\section{Fully Differential Distributions and Distinctions Between Interaction Structures}
\label{sec:QuadDiff}

Different Lagrangians for the HNL decay lead to different kinematics and hence measurements of the kinematic distributions may allow one to distinguish one new-physics scenario from all others. 
This may require a more detailed analysis beyond the single differential distribution, $d\Gamma/d\cos\theta_{\ell\ell}$,  which we have been concentrating on so far.  In this section, we investigate the ability of combinations of doubly-differential partial widths with respect to pairs of kinematical variables to distinguish interaction structures.  In particular, we assume that the outgoing charged leptons are nearly massless relative to the $N$, i.e.,  $m_{\alpha,\beta}\ll m_N $.  When considering MF decays and how they might differ from those of DF, we will make the further assumption that $\alpha = \beta$, \ie, the final-state charged leptons are identical.

\textbf{Dirac Fermion Cases:} Under the above assumptions there are four distinct ``types'' of decays for Dirac $N$ that we will define: pure scalar/pseudoscalar, pure vector/axial-vector (two types), and pure tensor. For the vector/axial-vector types, we take all $G_{NL}$ to be real, and find the two generic cases to be (I) $G_{VV} = G_{AA} = G_{AV} = G_{VA}$ and (II) $G_{VV} = -G_{AA} = -G_{AV} = G_{VA}$. In each different case the partial widths are
%
\begin{subnumcases}
{\label{eq:DiracAllCases}
\frac{d\Gamma(N \to \nu \ell_\alpha^- \ell_\beta^+)}{d z_{\ell\ell} dz_{\nu m} d \cos\theta_{\ell\ell} d \gamma_{\ell\ell} d \phi} =}
\mathcal{N} \left(p_p p_m \right) \left[ \left(p_\nu  p_N\right) + m_N \left(p_\nu s\right)\right],& \hspace{-0.6cm} (DF, SP) \label{eq:DiracSPcase}\\
\mathcal{N} \left(p_m p_\nu\right) \left[ \left(p_p p_N\right) + m_N \left(p_p s\right)\right],& \hspace{-0.6cm} (DF, VA I) \label{eq:DiracVAonecase} \\
\mathcal{N} \left(p_p p_\nu\right) \left[ \left(p_m p_N\right) - m_N \left(p_m s\right)\right],  & \hspace{-0.6cm} (DF, VA II) \label{eq:DiracVAtwocase} \\
\frac{\mathcal{N}}{3} \left[ 2\left(p_m p_\nu\right)\left(p_p p_N\right) + 2\left(p_p p_\nu\right)\left(p_m p_N\right) - \left(p_p p_m\right)\left(p_\nu p_N\right)\right].  & \hspace{-0.6cm} (DF, T)  \label{eq:DiracTcase}
\end{subnumcases}
The normalization factor is 
$\mathcal{N}=6\Gamma/(m_N^4\pi^2)$, $z_{\ell\ell} = m_{\ell\ell}^2/m_N^2$, and $z_{\nu m} = m_{\nu m}^2/m_N^2$.
For general vector/axial-vector interactions, we can have contributions from both ``types,'' and the associated differential distributions combine linearly. We have chosen the coefficients of the spin-dependent terms to maximize the forward/backward asymmetry. Moving away from this assumption will diminish the spin dependence and the resulting forward/backward asymmetry.

\textbf{Majorana Fermion Cases:} Fewer distinct cases arise for the MF case if we allow only two types of couplings at a time. The tensor case is identical to that in Eq.~\eqref{eq:DiracTcase}, so we do not repeat it here. If we have scalar/pseudoscalar couplings, then the spin-dependent piece necessarily vanishes. Finally, if we consider vector/axial-vector interactions, only one generic case exists, proportional to the sum of the ``VA I'' and ``VA II'' cases in Eq.~\eqref{eq:DiracAllCases}. The fully differential partial widths in the MF cases are
\begin{subnumcases}
{\label{eq:MajAllCases} \frac{d\Gamma
}{d\vec{\vartheta}}
=}
\mathcal{N} \left(p_p p_m \right) \left(p_\nu  p_N\right), & (MF,\ SP)\label{eq:MajSPcase}\\
\frac{\mathcal{N}}{2} \left[ \left(p_m p_\nu\right) \left(p_p p_N\right) + \left(p_p p_\nu\right) \left(p_m p_N\right) + m_N \left(\left(p_m p_\nu\right) \left(p_p s\right) - \left(p_p p_\nu\right) \left(p_m s\right)\right) \right]. & (MF, VA)\label{eq:MajVAcase}
\end{subnumcases}

\begin{figure}
\begin{center}
\includegraphics[width=0.48\linewidth]{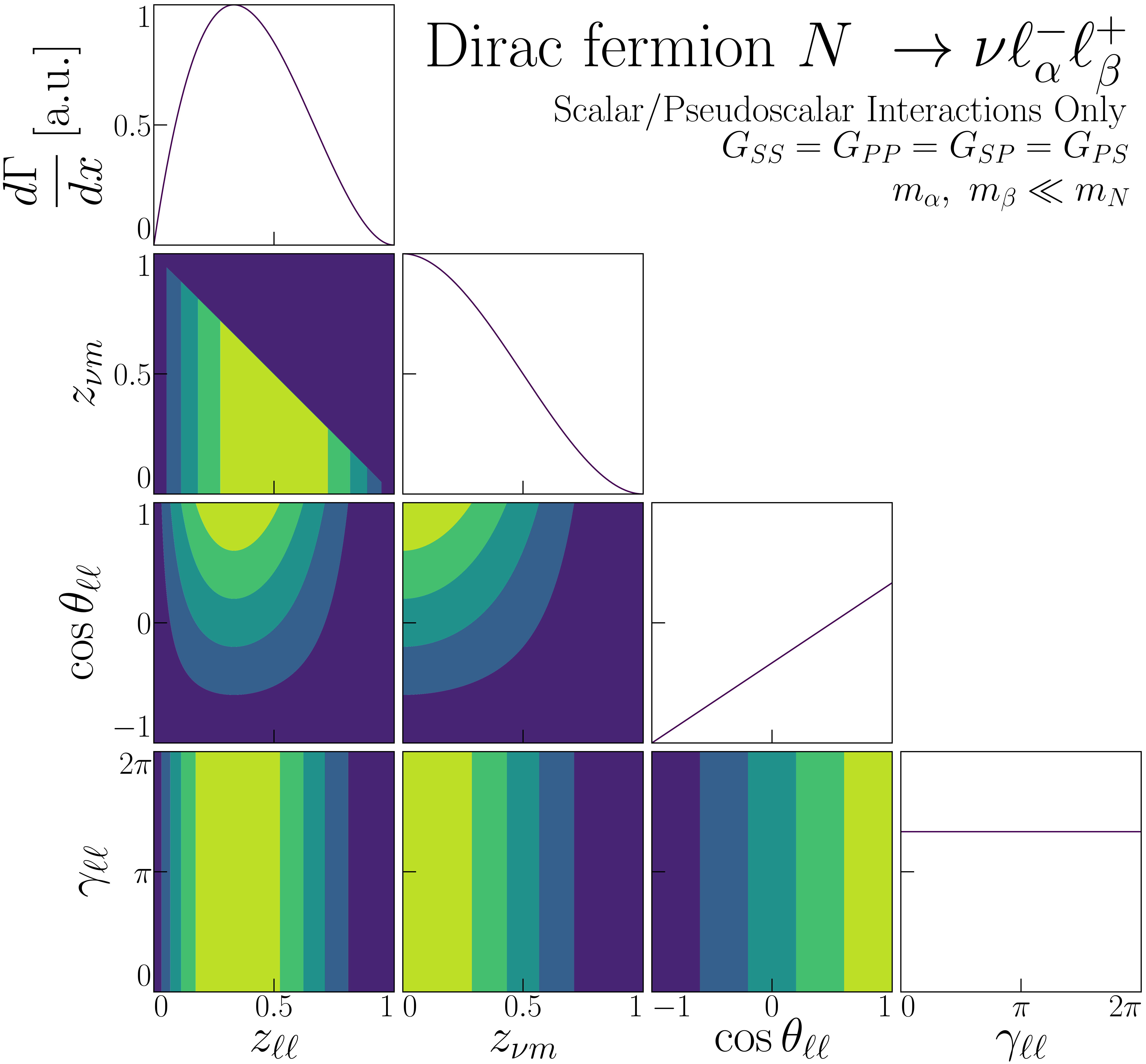}
\includegraphics[width=0.48\linewidth]{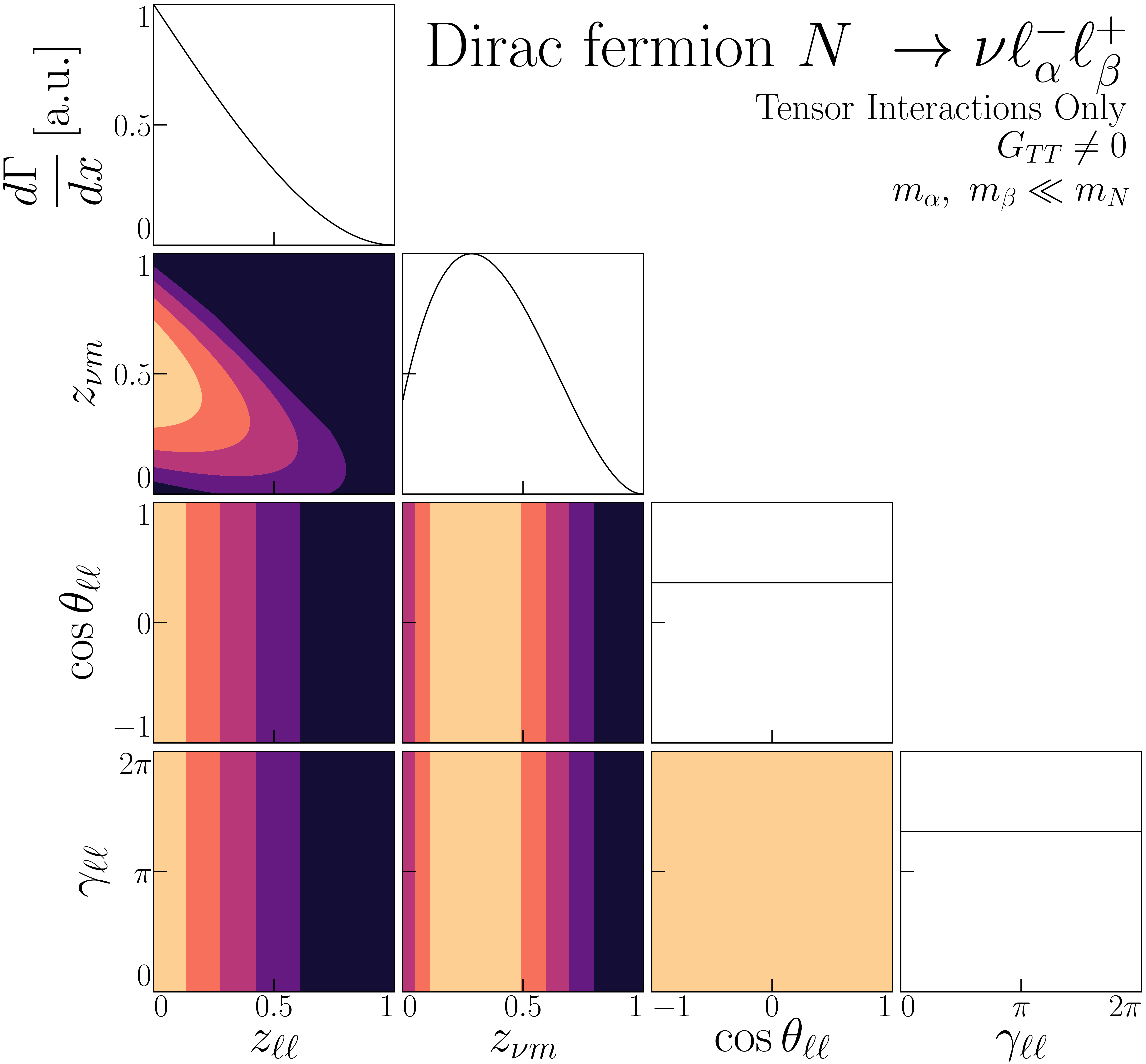} \\
\includegraphics[width=0.48\linewidth]{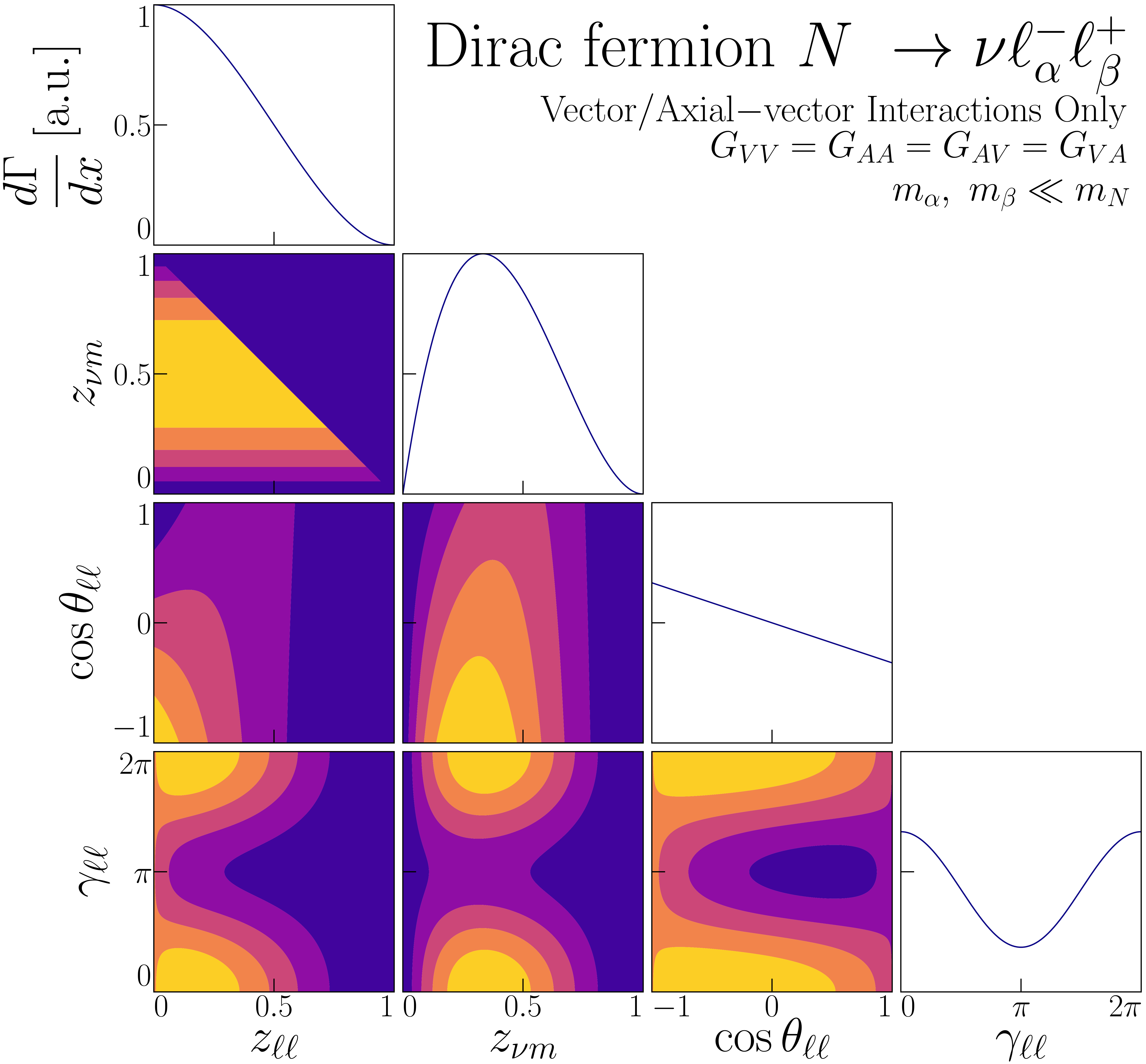}
\includegraphics[width=0.48\linewidth]{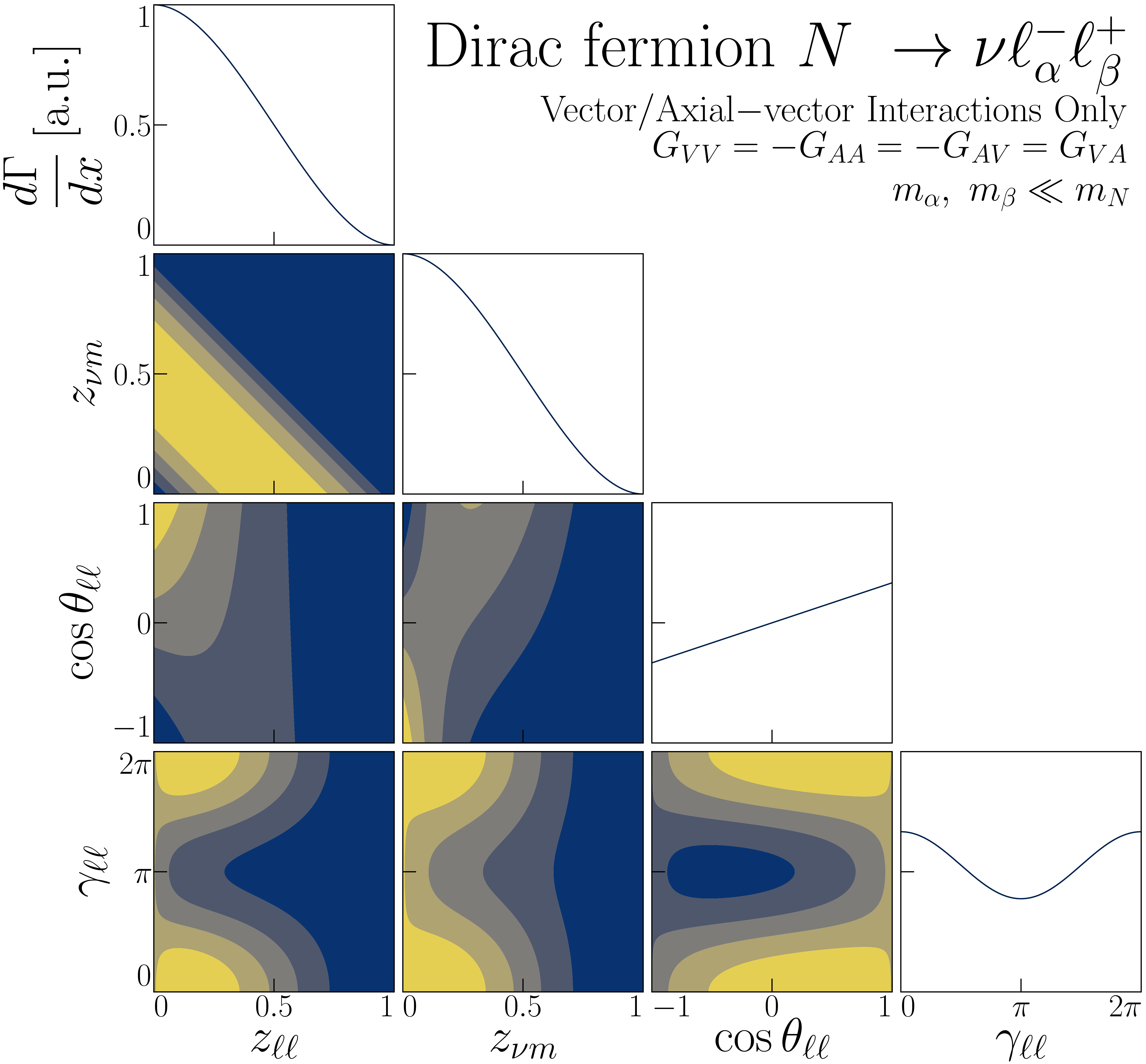}
\caption{Distributions of four different types of DF $N \to \nu \ell_\alpha^- \ell_\beta^+$ decays, in the limit $m_\alpha, m_\beta \ll m_N$. Each set contains, away from the diagonal sub-panels, two-dimensional distributions where the brighter (darker) colors correspond to larger (smaller) differential partial widths. One-dimensional distributions (diagonal sub-panels, solid lines) correspond to the single-differential partial widths $d\Gamma/dx$ with respect to each of the four kinematical variables $x \in \left\lbrace z_{\ell\ell}, z_{\nu m}, \cos\theta_{\ell\ell}, \gamma_{\ell\ell}\right\rbrace$, arbitrarily normalized.  For each of the cases presented, see Eqs.~\ref{eq:DiracAllCases}, the nonzero couplings are listed in each panel.
\label{fig:dGamma4PanelDirac}}
\end{center}
\end{figure}

Fig.~\ref{fig:dGamma4PanelDirac} depicts the doubly-differential distributions of DF $N$ decays for the four different types of decays. We show the doubly-differential distributions for each of the six pairs of final-state phase space parameters. Each two-dimensional panel has a color scale that is largest for light colors and smallest for dark ones. We also show the singly-differential partial widths along the diagonal for each of these, arbitrarily normalized.

We note several distinct features in the four different sets in Fig.~\ref{fig:dGamma4PanelDirac}. First, we see that in the one-dimensional $d\Gamma/d\cos\theta_{\ell\ell}$ panels, the slope is largest for the scalar/pseudoscalar case and zero for the tensor one -- this reflects the discussion in Section~\ref{subsec:AllowedAnisotropyDirac}: the allowed anisotropy can be large ($\Delta\Gamma/\Gamma = 1/2$) for scalar/pseudoscalar couplings. However, in the $m_\alpha,$ $m_\beta \ll m_N$ limit, for vector/axial-vector couplings, this anisotropy can only be as large as $\pm 1/6$. 
Different cases are also associated to different distributions in the  $z_{\ell\ell}$ vs. $z_{\nu m}$ panels in Fig.~\ref{fig:dGamma4PanelDirac}. For these four different cases, these distributions are qualitatively different. This implies that, with perfect measurements of these two parameters for a large-statistics sample, we could, in principle, distinguish between these different models for the DF HNL decay. 

\begin{figure}
\begin{center}
\includegraphics[width=0.48\linewidth]{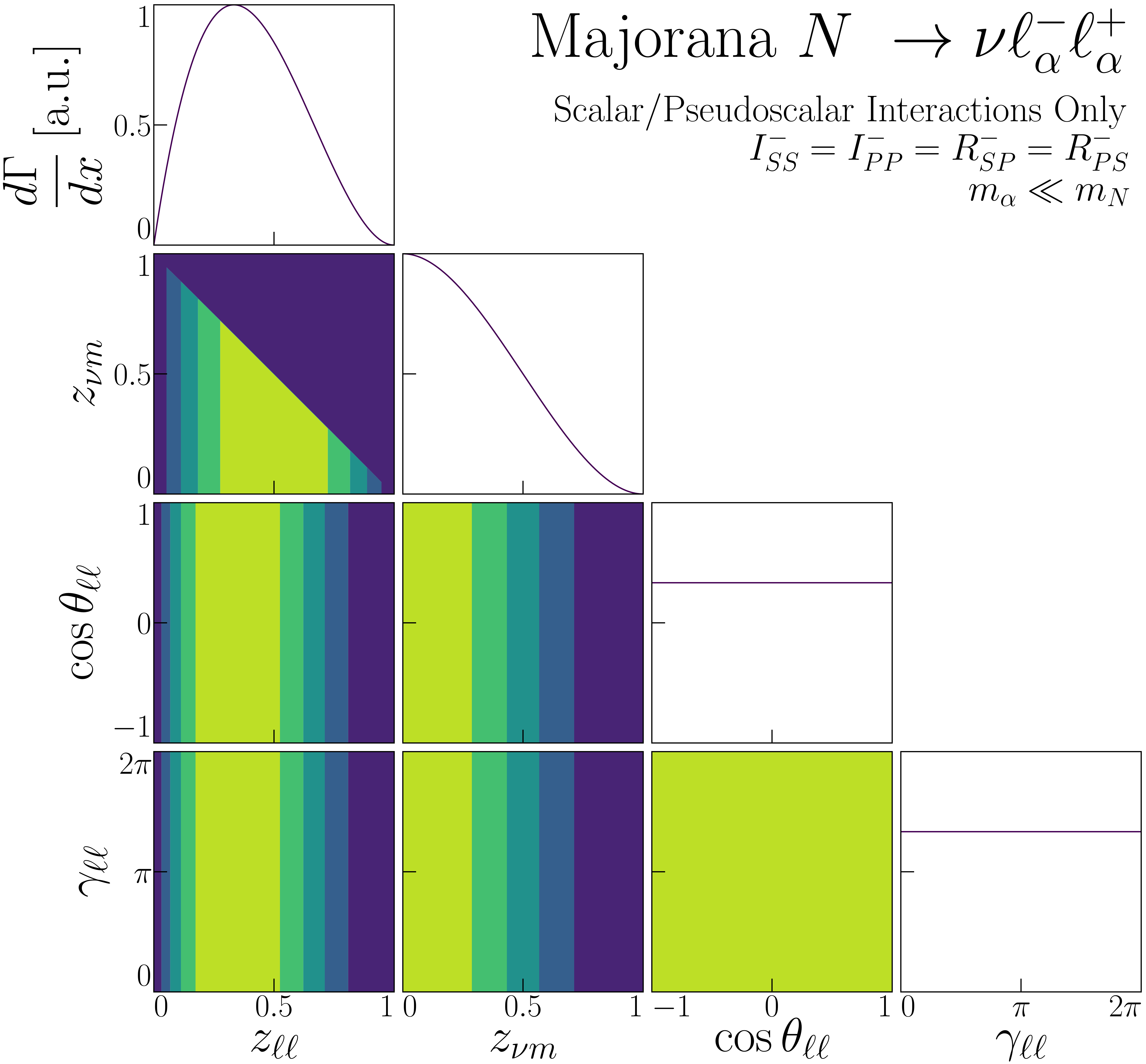} 
\includegraphics[width=0.48\linewidth]{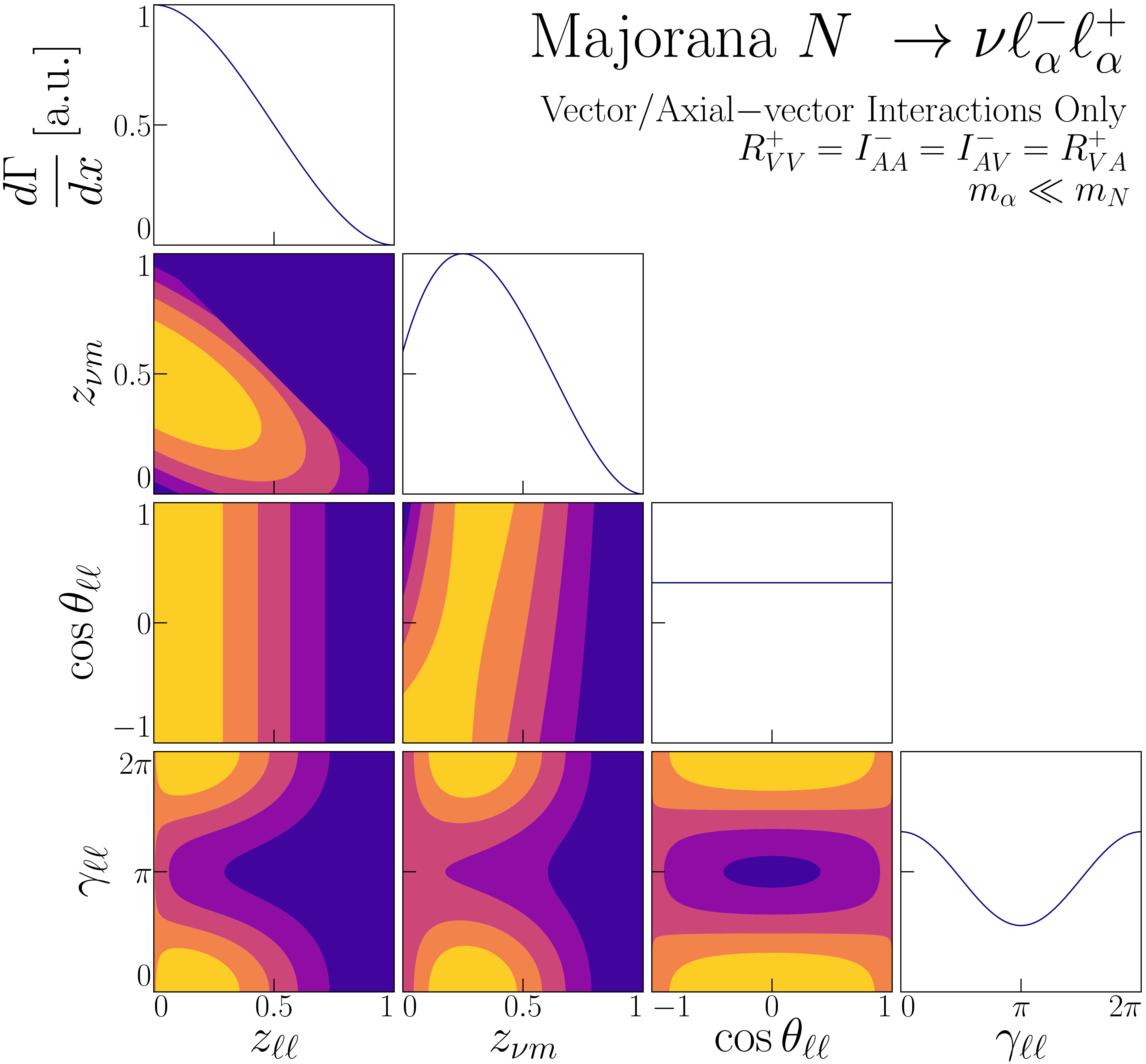}
\caption{Same as Fig.~\ref{fig:dGamma4PanelDirac}, but for two different MF $N$-decay models. Left: scalar and pseudoscalar interactions only, such that $R^-_{SP} = I^{-}_{SS} = R^{-}_{PS} = I^{-}_{PP}$. Right: vector/axial-vector interactions only, such that $R^+_{VV} = I^{-}_{AA} = R^{+}_{VA} = I^{-}_{AV}$.
\label{fig:dGamma2PanelMajorana}}
\end{center}
\end{figure}

Fig.~\ref{fig:dGamma2PanelMajorana} depicts the same distributions for MF $N$ decays. Contrasting Fig.~\ref{fig:dGamma2PanelMajorana} and Fig.~\ref{fig:dGamma4PanelDirac}, as expected, the one-dimensional distribution $d\Gamma/d\cos\theta_{\ell\ell}$ is flat\footnote{This result would hold even if we relaxed some of the assumptions in choosing sets of couplings -- as long as the two final-state charged leptons are identical then $d\Gamma/d\cos\theta_{\ell\ell}$ is flat.} for Majorana fermion $N$ decays. However, there is some parent-spin-dependence in the $\gamma_{\ell\ell}$ distribution in the vector/axial-vector case (right set of panels of Fig.~\ref{fig:dGamma2PanelMajorana}). This vanishes upon integrating over $\gamma_{\ell\ell} \in [0, 2\pi]$.

For MF $N$ decay, assuming vector/axial-vector interactions, the $z_{\ell\ell}$ vs. $z_{\nu m}$ panel takes on a different appearance than the two options in the Dirac fermion $N$ case: it is a linear combination of the two Dirac fermion options. Comparing the sets in Fig.~\ref{fig:dGamma2PanelMajorana} with those in Fig.~\ref{fig:dGamma4PanelDirac}, it appears that the $\cos\theta_{\ell\ell}$ dependence offers the strongest power to distinguish the MF and DF hypothesis using the kinematics of the $N$ three-body-decay.

\section{Discussion \& Conclusions}
\label{sec:Conclusions}
The fact that neutrinos have mass implies the existence of new particles or interactions beyond those that make up the standard model of particle physics. Whatever these might be, the neutrinos end up as either massive Dirac or massive Majorana fermions. This distinction is fundamentally linked to whether lepton-number symmetry is conserved (Dirac) or violated (Majorana) in nature.

Considerable research has been dedicated to ways of observing lepton-number violation in the laboratory in order to shed light on this question. These searches operate on the principle that if lepton-number violation is observed, then the Majorana-fermion nature of neutrinos is confirmed. On the contrary, few strategies exist to confirm whether neutrinos are Dirac fermions. Recently, the strategy of observing distributions of two-body decays of neutral fermions was highlighted as a means of confirming the Dirac-fermion nature of neutrinos.

In this work, we expanded upon this idea by studying three-body decays of Dirac and Majorana fermions in great detail. We focused on the case where a heavy fermion, such as a heavy neutral lepton ($N$) that mixes with the light neutrinos, decays into a light neutrino and a pair of charged leptons.  We studied $N$-decays through generic contact interactions, considering all possible combinations of four-fermion interactions, defined in Eq.~\eqref{eq:Lagrangian}.  Furthermore, we considered final states containing both like-flavor and different-flavor charged leptons, and detectors that can and cannot distinguish the charge of the final state leptons.  We determined several features that allow for distinction between the three-body decays of Dirac and Majorana fermions.  Under the restriction of like-flavor charged leptons or a charge-blind detector, we used CPT arguments to show that Majorana-fermion decays exhibit isotropy (leading to $A_{\rm FB} = 0$) in the direction of the outgoing neutrino (or equivalently, the outgoing charged-lepton pair), whereas Dirac fermion decays can have large anisotropies.

Going beyond studying forward/backward asymmetries, we also explored the fully-differential phase space of the three-body decays.  We demonstrated how to express the fully differential partial width for $N$ decaying in its rest frame Eq.~\eqref{eq:dGamma} in terms of $13$ kinematic invariants Eqs.~\eqref{eq:K1}-\eqref{eq:K13}.  In extensive appendices we tabulate the contributions of each kinematic invariant to the partial width, under various assumptions for the Dirac/Majorana nature of $N$, the flavor structure of the final state, and the capabilities of the detector, as well as provide the technical details of the calculations.  These results allow us to confirm the conclusions of the simple CPT arguments and to move beyond the assumptions associated with them to study more general cases.  
We also demonstrated how these full distributions provide more leverage in distinguishing between Dirac and Majorana fermion decays, and also can help to determine the structure of the interactions mediating the decays.

As an example of the utility of the general results, we showed that, for Dirac fermions, if \emph{only} two types of coupling are nonzero, anisotropy can occur for only scalar/pseudoscalar couplings ($|A_{\rm FB}| \le 1/2$) or only vector/axial-vector couplings ($|A_{\rm FB}| \le 1/6$ for heavy $N$ and $|A_{\rm FB}| \le 1/2$ if final state lepton masses cannot be ignored), while all other combinations of pairs of couplings result in $A_{\rm FB}=0$.  The situation is more complicated if more than two types of couplings are nonzero. The case of Majorana fermions decaying to different-flavor charged-leptons  through neutral mediators only requires at least three different types of coupling to generate a forward/backward asymmetry, and only with all couplings non-zero and $N$ light can the asymmetry be as large as for a Dirac fermion with scalar/pseudoscalar couplings, as depicted Fig.~\ref{fig:AlphaMaj}.  This key distinction, that Majorana-fermion decays tend to have zero forward/backward asymmetry in the direction of the outgoing neutrino, means that it is often possible to distinguish between the Majorana and Dirac fermion hypotheses using kinematic distributions, see Figs.~\ref{fig:dGamma4PanelDirac} and \ref{fig:dGamma2PanelMajorana}.

Throughout this work, we have restricted ourselves to analytic calculations of these decays in the rest frame of the decaying particle, speculating on differences between distributions that can be leveraged once a heavy fermion is hypothetically discovered. In our companion paper \cite{deGouvea:2021TBA}, under preparation, we will take this framework and apply it to several phenomenological cases of interest, determining the required statistics to distinguish between different model scenarios.

As we venture forward attempting to discover the nature of the neutrinos, the question of lepton-number conservation is crucial. We demonstrated here that, in tandem with other experimental searches for lepton-number violation, the decays (specifically, three-body decays) of heavy neutrinos are a new tool that might be leveraged to address this fundamental puzzle.

\section*{Acknowledgements}
The work of AdG is supported in part by the DOE Office of Science
award \#DE-SC0010143.
PF, BK, and KJK are supported by Fermi Research Alliance, LLC under Contract DE-AC02-07CH11359 with the U.S. Department of Energy. AdG and KJK thank the Institute for Nuclear Theory at the University of Washington for its hospitality during which a portion of this work was completed.

\appendix
\clearpage
\newpage

\section{Matrix-Elements-Squared for Heavy Neutral Lepton Decays}\label{appendix:MSqs}\setcounter{footnote}{0}

In Section~\ref{subsec:LIs}, we introduced the set of thirteen Lorentz Invariants $K_i$ for a convenient decomposition of the $N$ decay matrix-element-squared. In this language, $\left\lvert \mathcal{M}\right\rvert^2 = \sum_i C_i K_i$, where $C_i$ are coefficients depending on the $G_{NL}$ and/or $\overline{G}_{NL}$ entering the matrix elements of interest. In this appendix, we provide the full expressions for $C_i$ in different scenarios, specifically
\begin{itemize}
\item Dirac fermion $N$ with contributions from $\mathcal{M}_1$. This is given in Table~\ref{table:MSqDirac}.
\item Dirac fermion $\overline{N}$ with contributions from $\mathcal{M}_2$. This is given in Table~\ref{table:MSqDiracBar}.
\item Majorana fermion $N$ with contributions from $\mathcal{M}_1$ and $\mathcal{M}_2$.
\begin{itemize}
\item Generic result with no mediator assumptions, given in Table~\ref{table:MSqMajGen}.
\item Results assuming only neutral mediators contribute, given in Table~\ref{table:MSqMajNMO}.
\end{itemize}
\item Majorana fermion $N$ with identical final-state charged leptons, with contributions from $\mathcal{M}_1$, $\mathcal{M}_2$, $\mathcal{M}_1^c$, and $\mathcal{M}_2^c$. 
\begin{itemize}
\item Results obtained without mediator assumptions provided in Table~\ref{table:MSqMajAB}.
\item Subsequent results with only neutral mediators are given in Table~\ref{table:MSqMajABNMO}.
\end{itemize}
\end{itemize}

In order to simplify our results for the various Majorana fermion calculations, we define some new parameters:
\begin{align}
G^{\pm}_{NL} &\equiv \left(G_{NL} \pm \overline{G}_{NL}\right),\\
R_{NL}^{\pm} &\equiv \mathrm{Re}\left(G_{NL} \pm \overline{G}_{NL}\right),\\
I_{NL}^{\pm} &\equiv \mathrm{Im}\left(G_{NL} \pm \overline{G}_{NL}\right).
\end{align}
We find that the $C_i$ for a Majorana fermion only depend on either $G^+_{NL}$ \textit{or} $G^-_{NL}$ for any given combination of $(NL)$. Additionally, when we make further restrictions, such as assuming the final-state charged leptons are identical, only dependence on one of $R^+_{NL}$, $R^-_{NL}$, $I^+_{NL}$, $I^-_{NL}$ appears for any given $(NL)$. We express our results in terms of $G_{NL}^\pm$, $R_{NL}^\pm$, and $I_{NL}^\pm$ for our Majorana fermion results in Tables~\ref{table:MSqMajGen} and~\ref{table:MSqMajAB}.

The results of the Majorana Fermion $N$ decay calculations given in Tables~\ref{table:MSqMajGen} and~\ref{table:MSqMajAB} can equivalently be determined by using the results given for Dirac fermions with specific substitutions. Specifically, the general Majorana fermion $N$ decay result (Table~\ref{table:MSqMajGen}) is equivalent to taking the Dirac fermion $N$ result (Table~\ref{table:MSqDirac}) with the following set of replacements:
\begin{align}\label{eq:appDirtoMajstart}
&G_{SS} \to G_{SS} - \overline{G}_{SS} = G^-_{SS},\quad\quad G_{SP} \to G_{SP} - \overline{G}_{SP} = G^-_{SP}, \\
&G_{PS} \to G_{PS} - \overline{G}_{PS} = G^-_{PS},\quad\quad G_{PP} \to G_{PP} - \overline{G}_{PP} = G^-_{PP}, \\
&G_{VV} \to G_{VV} + \overline{G}_{VV} = G^+_{VV},\quad\quad G_{AV} \to G_{AV} - \overline{G}_{AV} = G^-_{AV}, \\
&G_{VA} \to G_{VA} + \overline{G}_{VA} = G^+_{VA},\quad\quad G_{AA} \to G_{AA} - \overline{G}_{AA} = G^-_{AA}, \\
&G_{TT} \to G_{TT} + \overline{G}_{TT} = G^+_{TT}.
\label{eq:appDirtoMajend}
\end{align}
This demonstrates why, for each $(NL)$, only one coupling of the type $G^\pm_{NL}$ appears in Table~\ref{table:MSqMajGen}. In order to obtain the results of Table~\ref{table:MSqMajAB}, when the final-state charged leptons are identical, we may perform a set of substitutions on Table~\ref{table:MSqMajGen}. These are
\begin{align}
&G^-_{SS} \to 2I^-_{SS},\quad G^-_{SP} \to 2R^-_{SP},\quad G^-_{PS} \to 2R^-_{PS},\quad G^-_{SS}\to 2I^-_{SS}, \\
&G^+_{VV} \to 2R^+_{VV},\quad G^-_{AV} \to 2I^-_{AV},\quad G^+_{VA}\to 2R^+_{VA},\quad G^-_{AA}\to 2I^-_{AA}, \\
&G^+_{TT} \to 2R^+_{TT}.
\end{align}

These results can be used to determine the Lorentz-invariant coefficients for Dirac fermion $N$ and $\overline{N}$ if the final-state charged leptons are identical. These are not given in any table due to their cumbersome forms. However, they can be obtained by substitution on Tables~\ref{table:MSqDirac} and~\ref{table:MSqDiracBar}, respectively. The substitutions for Dirac fermion $N$ are
\begin{equation}
G_{NL} \longrightarrow G_{NL} + \eta_{N}\eta_{L}\overline{G}_{NL}^*, \label{eq:DiracNAB}
\end{equation}
where $\eta_{N,L}$ previously appeared in our definitions of $\mathcal{M}_1^c$ and $\mathcal{M}_2^c$ in Eqs.~\eqref{eq:M1c} and~\eqref{eq:M2c} -- a reminder that $\eta_X = +1$ for $X = S, V, A, T$ and $\eta_X = -1$ for $N = X$. For Dirac fermion $\overline{N}$, the appropriate substitutions are
\begin{equation}
\overline{G}_{NL} \longrightarrow \overline{G}_{NL} + \eta_{N}\eta_{L} G_{NL}^*.\label{eq:DiracNBarAB}
\end{equation}
When we are considering Dirac fermion $N$ or $\overline{N}$ decays of the type $N \to \nu \ell_\alpha^+ \ell_\alpha^-$, then the two decays ($N$ and $\overline{N}$) must yield the same partial width at tree level. Under the substitutions of Eqs.~\eqref{eq:DiracNAB} and~\eqref{eq:DiracNBarAB}, the coefficients $C_1$ through $C_6$ (which contribute to the total width of the decay) are not exactly identical, however their differences cancel out when considering the total width of $N$ and $\overline{N}$. Likewise, the coefficients $C_7$ through $C_{13}$, which can provide an overall forward-backward asymmetry, are related in a way that the two distributions, $N$ and $\overline{N}$ necessarily have equal and opposite forward-backward asymmetries.

Inspecting Table~\ref{table:MSqMajAB}, we see that certain $C_i$ are related, e.g. $C_2 = C_3$, and others are necessarily zero: $C_7 = C_8 = 0$. This, along with the relations $C_{10} = -C_9$ and $C_{12} = -C_{11}$ leads to our result that, when decaying to identical final-state charged leptons, Majorana fermions have zero forward-backward asymmetry. In order to connect this relationship of $C_i$ to a zero forward-backward asymmetry, we must integrate the Lorentz invariants over a subset of phase space -- this integration is performed in Appendix~\ref{app:Integration}.

Before concluding this appendix, we will focus on one final sub-class of models. This is when all mediators of $N$ and $\overline{N}$ decay are neutral (or at least, overwhelmingly dominant in the decay width contributions). This allows us to apply further restrictions to the coefficients $G_{NL}$ and $\overline{G}_{NL}$ and arrive at further simplifications to the $C_i$ obtained in this appendix.

\subsection{Restrictions when Mediators are Neutral}\label{appendix:NMO}
In writing the matrix elements of Eqs.~\eqref{eq:MN} and~\eqref{eq:MNBar}, we chose to express them as a four-fermion terms in the ``neutral'' ordering, contracting the spinors of $N$ and $\nu$, and the charged leptons with each other. Because we allowed for a generic set of $\Gamma_{N}$ and $\Gamma_{L}$, this choice was generic, even if the mediators are charged, due to Fierz rearrangement. If the new mediators are all neutral, then this ordering is even more well-motivated, and any term in the matrix-element $G_{NL}$ can be thought of as coming from a product of two fundamental couplings -- one coupling of the new mediator with the $N\nu$ vertex and one with the charged lepton vertex. We introduce a new notation here,
\begin{equation}
G_{NL} \longrightarrow \frac{1}{m_\phi^2} g_{\nu N} g_{\ell L} \quad \quad \mathrm{(neutral\ mediators\ only)}, \label{eq:NMO}
\end{equation}
where $m_\phi$ is the mass scale of the new mediator(s) and $g_{\nu N}$ and $g_{\ell L}$ are dimensionless couplings.

In this case, the couplings $G_{NL}$ and $\overline{G}_{NL}$ are related by charge conjugation. Specifically,
\begin{equation}
\overline{G}_{NL} \longrightarrow \eta_N \frac{1}{m_\phi^2} g_{\nu N}^* g_{\ell L}~. \label{eq:NMOBar}
\end{equation}
As a consequence, many relations between various $G_{NL}$ and $\overline{G}_{N^\prime L^\prime}$ may be derived, e.g. $\left\lvert G_{NL} \right\rvert = \left \lvert \overline{G}_{NL}\right\rvert$, among others. These lead to significant simplification of the coefficients we obtained in Table~\ref{table:MSqMajGen} if $N$ is a Majorana fermion with decays mediated only by neutral mediators.

As a specific example, let us explore some terms that appear in $C_7$ and $C_8$ of Table~\ref{table:MSqMajGen}, including
\begin{equation}
C_7^{\rm Maj.} = 16 \mathrm{Re}\left[ G_{PP}^- (G^-_{SP})^* + G^-_{SS} (G^-_{PS})^*\right].
\end{equation}
Under the replacements of Eqs.~\eqref{eq:NMO} and~\eqref{eq:NMOBar}, this becomes
\begin{align}
C_7^{\rm Maj., NMO} &\to 16 \mathrm{Re} \left[ \left(g_{\nu P} g_{\ell P} + g_{\nu P}^* g_{\ell P}\right) \left( g_{\nu S}^* g_{\ell P}^* - g_{\nu S} g_{\ell P}^*\right)  + \left( g_{\nu S} g_{\ell S} - g_{\nu S}^* g_{\ell S}\right) \left( g_{\nu P}^* g_{\ell S}^* + g_{\nu P} g_{\ell S}^*\right) \right], \\
&= 16\mathrm{Re} \left[ -4i\absq{g_{\ell P}} \mathrm{Re}\left(g_{\nu P}\right) \mathrm{Im}\left(g_{\nu S}\right) + 4i\absq{g_{\ell S}} \mathrm{Im}\left(g_{\nu S}\right) \mathrm{Re}\left(g_{\nu P}\right) \right],\label{eq:C7Rep}\\
&= 0.
\end{align}
In replacements of this type, we also find that, because Majorana fermion decays only depended on $G^\pm_{NL}$, only the real or imaginary part of each $g_{\nu N}$ survives for different $N$. As we see in Eq.~\eqref{eq:C7Rep}, only the real (imaginary) part of $g_{\nu P}$ ($g_{\nu S}$) appears. We will take this dependence into account in our results and perform some minor notational substitutions, $\mathrm{Re}(g_{\nu P}) \to g_{\nu P}$, $\mathrm{Im}(g_{\nu S}) \to g_{\nu S}$, etc., to simplify our results.

We provide the full results for Majorana fermion decays assuming neutral mediators only in Table~\ref{table:MSqMajNMO}. Such a table could also be derived for Dirac fermion $N$ and $\overline{N}$ decays, however, the results are still as complicated as in Tables~\ref{table:MSqDirac} and~\ref{table:MSqDiracBar}, so we omit them for brevity.

If we assume that only neutral mediators contribute to $N$ decay \textit{and} that the final-state charged leptons are identical, then further simplifications occur. We find that only the real/imaginary parts of specific $g_{\ell L}$ (like with the various $g_{\nu N}$ above) survive, and perform similar simplifying replacements. We find even more cancellations in this case to the point that the resulting $C_i$ are simpler and have even further restrictions: $C_2 = C_3$, $C_4 = C_5$, $C_7 = C_8 = 0$, $C_9 = -C_{10}$, and $C_{11} = - C_{12}$. The results in this case are presented in Table~\ref{table:MSqMajABNMO}.

\begin{table}[!h]
\begin{center}
\caption{Lorentz-invariant decomposition coefficients for a DF $N$ decaying via the matrix element in Eq.~\eqref{eq:MN}. The corresponding Lorentz-invariant quantities $K_i$ are given in Eqs.~(\ref{eq:K1})-(\ref{eq:K13}), and the matrix-element squared can be expressed as in Eq.~\eqref{eq:MSqDecomposed}.\label{table:MSqDirac}}
\begin{tabular}{c|c}\hline\hline
$C_i$ & Dirac Fermion $N$  \\ \hline
$C_1$ & $8\left( \absq{G_{PP}} + \absq{G_{SP}} - \absq{G_{PS}} - \absq{G_{SS}}\right) + 16\left( \absq{G_{VV}} + \absq{G_{AV}} - \absq{G_{VA}} - \absq{G_{AA}}\right)$ \\ 
$C_2$ & $-96\mathrm{Re}\left[\left(G_{AA}-G_{VV}\right)G_{TT}^*\right] - 16\mathrm{Re}\left[ G_{AA} G_{PP}^* + G_{AV}G_{PS}^* - G_{VA} G_{SP}^* - G_{VV} G_{SS}^*\right]$ \\ 
$C_3$ & $96\mathrm{Re}\left[\left(G_{AA}+G_{VV}\right)G_{TT}^*\right] - 16\mathrm{Re}\left[G_{AA} G_{PP}^* - G_{AV} G_{PS}^* - G_{VA} G_{SP}^* + G_{VV} G_{SS}^*\right]$ \\ 
$C_4$ & $16\left[ \absq{G_{AA}+G_{VV}} + \absq{G_{VA}+G_{AV}}\right] + 128\absq{G_{TT}} - 32 \mathrm{Re}\left[\left(G_{SS}+G_{PP}\right)G_{TT}^*\right]$ \\ 
$C_5$ & $16\left[ \absq{G_{AA} - G_{VV}} + \absq{G_{VA} - G_{AV}}\right] + 128 \absq{G_{TT}} +32 \mathrm{Re}\left[\left(G_{SS}+G_{PP}\right)G_{TT}^*\right]$ \\ 
$C_6$ & $8\left[\absq{G_{PP}} + \absq{G_{SP}} + \absq{G_{PS}} + \absq{G_{SS}}\right] - 64 \absq{G_{TT}}$ \\ \hline
$C_7$ & $16\mathrm{Re}\left[ G_{PP} G_{SP}^* + G_{SS} G_{PS}^*\right]$ \\ 
$C_8$ & $16\mathrm{Re}\left[ G_{PP} G_{SP}^* - G_{SS} G_{PS}^*\right] - 32\mathrm{Re}\left[ G_{AA} G_{VA}^* - G_{VV} G_{AV}^*\right]$ \\ 
$C_9$ & $32\mathrm{Re}\left[\left(G_{VV}+G_{AA}\right)\left(G_{AV}^*+G_{VA}^*\right)\right] - 32\mathrm{Re}\left[\left(G_{SP}+G_{PS}\right)G_{TT}^*\right]$ \\ 
$C_{10}$ & $32\mathrm{Re}\left[\left(G_{VV}-G_{AA}\right)\left(G_{AV}^*-G_{VA}^*\right)\right] + 32\mathrm{Re}\left[\left(G_{SP}+G_{PS}\right)G_{TT}^*\right]$ \\ 
$C_{11}$ & $16\mathrm{Re}\left[G_{AA} G_{SP}^* + G_{AV} G_{SS}^* - G_{VA} G_{PP}^* - G_{VV} G_{PS}^*\right] - 32 \mathrm{Re}\left[\left(G_{AV}-G_{VA}\right)G_{TT}^*\right]$ \\ 
$C_{12}$ & $16\mathrm{Re}\left[G_{AA} G_{SP}^* - G_{AV} G_{SS}^* - G_{VA} G_{PP}^* + G_{VV} G_{PS}^*\right] - 32 \mathrm{Re}\left[\left(G_{AV}+G_{VA}\right)G_{TT}^*\right]$ \\ 
$C_{13}$ & $-32 m_N\mathrm{Im}\left[\left(G_{PP}+G_{SS}\right)G_{TT}^*\right]$ \\ 
& $+16m_\alpha \left( \mathrm{Im}\left[ G_{PP} G_{AA}^* + G_{PS} G_{AV}^* - G_{SP} G_{VA}^* - G_{SS}G_{VV}^*\right]  -2 \mathrm{Im}\left[\left(G_{VV}-G_{AA}\right)G_{TT}^*\right]\right)$ \\
& $+16m_\beta \left( \mathrm{Im}\left[ -G_{PP} G_{AA}^* + G_{PS} G_{AV}^* + G_{SP} G_{VA}^* - G_{SS}G_{VV}^*\right] + 2\mathrm{Im}\left[\left(G_{AA}+G_{VV}\right)G_{TT}^*\right]\right)$ \\\hline\hline
\end{tabular}
\end{center}
\end{table}

\begin{table}[!h]
\begin{center}
\caption{Identical to Table~\ref{table:MSqDirac} but for a DF $\overline{N}$, decaying via the matrix element in Eq.~\eqref{eq:MNBar}.
\label{table:MSqDiracBar}}
\small
\begin{tabular}{c|c}\hline\hline
$C_i$ & Dirac Fermion $\overline{N}$  \\ \hline
$C_1$ & $8\left( \absq{\overline{G}_{PP}} + \absq{\overline{G}_{SP}} - \absq{\overline{G}_{PS}} - \absq{\overline{G}_{SS}}\right) + 16\left( \absq{\overline{G}_{VV}} + \absq{\overline{G}_{AV}} - \absq{\overline{G}_{VA}} - \absq{\overline{G}_{AA}}\right)$ \\ 
$C_2$ & $96\mathrm{Re}\left[\left(\overline{G}_{AA}+\overline{G}_{VV}\right)\overline{G}_{TT}^*\right] - 16\mathrm{Re}\left[ \overline{G}_{AA} \overline{G}_{PP}^* + \overline{G}_{AV}\overline{G}_{PS}^* + \overline{G}_{VA} \overline{G}_{SP}^* + \overline{G}_{VV} \overline{G}_{SS}^*\right]$ \\ 
$C_3$ & $-96\mathrm{Re}\left[\left(\overline{G}_{AA}-\overline{G}_{VV}\right)\overline{G}_{TT}^*\right] - 16\mathrm{Re}\left[\overline{G}_{AA} \overline{G}_{PP}^* - \overline{G}_{AV} \overline{G}_{PS}^* + \overline{G}_{VA} \overline{G}_{SP}^* - \overline{G}_{VV} \overline{G}_{SS}^*\right]$ \\ 
$C_4$ & $16\left[ \absq{\overline{G}_{AA}-\overline{G}_{VV}} + \absq{\overline{G}_{VA}-\overline{G}_{AV}}\right] + 128\absq{\overline{G}_{TT}} + 32 \mathrm{Re}\left[\left(\overline{G}_{SS}+\overline{G}_{PP}\right)\overline{G}_{TT}^*\right]$ \\ 
$C_5$ & $16\left[ \absq{\overline{G}_{AA} + \overline{G}_{VV}} + \absq{\overline{G}_{VA} + \overline{G}_{AV}}\right] + 128\absq{\overline{G}_{TT}} - 32\mathrm{Re}\left[\left(\overline{G}_{SS}+\overline{G}_{PP}\right)\overline{G}_{TT}^*\right]$ \\ 
$C_6$ & $8\left[\absq{\overline{G}_{PP}} + \absq{\overline{G}_{SP}} + \absq{\overline{G}_{PS}} + \absq{\overline{G}_{SS}}\right] - 64\absq{\overline{G}_{TT}}$ \\ \hline
$C_7$ & $16\mathrm{Re}\left[ \overline{G}_{PP} \overline{G}_{SP}^* + \overline{G}_{SS} \overline{G}_{PS}^*\right]$ \\ 
$C_8$ & $16\mathrm{Re}\left[ \overline{G}_{PP} \overline{G}_{SP}^* - \overline{G}_{SS} \overline{G}_{PS}^*\right] + 32\mathrm{Re}\left[ \overline{G}_{AA} \overline{G}_{VA}^* - \overline{G}_{VV} \overline{G}_{AV}^*\right]$ \\ 
$C_9$ & $32\mathrm{Re}\left[\left(\overline{G}_{VV}-\overline{G}_{AA}\right)\left(\overline{G}_{VA}^*-\overline{G}_{AV}^*\right)\right] + 32 \mathrm{Re}\left[\left(\overline{G}_{SP}+\overline{G}_{PS}\right)\overline{G}_{TT}^*\right]$ \\ 
$C_{10}$ & $-32\mathrm{Re}\left[\left(\overline{G}_{VV}+\overline{G}_{AA}\right)\left(\overline{G}_{VA}^*+\overline{G}_{AV}^*\right)\right] - 32 \mathrm{Re}\left[\left(\overline{G}_{SP}+\overline{G}_{PS}\right)\overline{G}_{TT}^*\right]$ \\ 
$C_{11}$ & $16\mathrm{Re}\left[\overline{G}_{AA} \overline{G}_{SP}^* + \overline{G}_{AV} \overline{G}_{SS}^* + \overline{G}_{VA} \overline{G}_{PP}^* + \overline{G}_{VV} \overline{G}_{PS}^*\right] + 32 \mathrm{Re}\left[\left(\overline{G}_{AV}+\overline{G}_{VA}\right)\overline{G}_{TT}^*\right]$ \\ 
$C_{12}$ & $16\mathrm{Re}\left[\overline{G}_{AA} \overline{G}_{SP}^* - \overline{G}_{AV} \overline{G}_{SS}^* + \overline{G}_{VA} \overline{G}_{PP}^* - \overline{G}_{VV} \overline{G}_{PS}^*\right] + 32 \mathrm{Re}\left[\left(\overline{G}_{AV}-\overline{G}_{VA}\right)\overline{G}_{TT}^*\right]$ \\ 
$C_{13}$ & $ 32 m_N\mathrm{Im}\left[\left(\overline{G}_{PP}+\overline{G}_{SS}\right)\overline{G}_{TT}^*\right]$ \\ 
& $+16m_\alpha \left( \mathrm{Im}\left[ \overline{G}_{PP} \overline{G}_{AA}^* + \overline{G}_{PS} \overline{G}_{AV}^* + \overline{G}_{SP} \overline{G}_{VA}^* + \overline{G}_{SS}\overline{G}_{VV}^*\right] -2 \mathrm{Im}\left[\left(\overline{G}_{AA}+\overline{G}_{VV}\right)\overline{G}_{TT}^*\right]\right)$ \\
& $+16m_\beta \left( \mathrm{Im}\left[ -\overline{G}_{PP} \overline{G}_{AA}^* + \overline{G}_{PS} \overline{G}_{AV}^* - \overline{G}_{SP} \overline{G}_{VA}^* + \overline{G}_{SS}\overline{G}_{VV}^*\right] +2 \mathrm{Im}\left[\left(\overline{G}_{VV}-\overline{G}_{AA}\right)\overline{G}_{TT}^*\right]\right)$ \\\hline\hline
\end{tabular}
\end{center}
\end{table}

\begin{table}[!h]
\begin{center}
\caption{Identical to Table~\ref{table:MSqDirac} but for a MF $N$ decaying via the matrix elements in Eqs.~\eqref{eq:MN} and~\eqref{eq:MNBar}.\label{table:MSqMajGen}}
\small
\begin{tabular}{c|c}\hline\hline
$C_i$ & Majorana Fermion $N$  \\ \hline
$C_1$ & $8\left( \absq{G_{PP}^{-}} + \absq{G_{SP}^{-}} - \absq{G_{PS}^{-}} - \absq{G_{SS}^{-}}\right) + 16\left( \absq{G_{VV}^{+}} + \absq{G_{AV}^{-}} - \absq{G_{VA}^{+}} - \absq{G_{AA}^{-}}\right)$ \\ 
$C_2$ & $-96\mathrm{Re}\left[\left(G_{AA}^{-}-G_{VV}^{+}\right)(G_{TT}^{+})^*\right] - 16\mathrm{Re}\left[ G_{AA}^{-} (G_{PP}^{-})^* + G_{AV}^{-}(G_{PS}^{-})^* - G_{VA}^{+} (G_{SP}^{-})^* - G_{VV}^{+} (G_{SS}^{-})^*\right]$ \\ 
$C_3$ & $96\mathrm{Re}\left[\left(G_{AA}^{-}+G_{VV}^{+}\right)(G_{TT}^{+})^*\right] - 16\mathrm{Re}\left[G_{AA}^{-} (G_{PP}^{-})^* - G_{AV}^{-} (G_{PS}^{-})^* - G_{VA}^{+} (G_{SP}^{-})^* + G_{VV}^{+} (G_{SS}^{-})^*\right]$ \\ 
$C_4$ & $16\left[ \absq{G_{AA}^{-}+G_{VV}^{+}} + \absq{G_{VA}^{+}+G_{AV}^{-}}\right] + 128\absq{G_{TT}^{+}} - 32 \mathrm{Re}\left[\left(G_{SS}^{-}+G_{PP}^{-}\right)(G_{TT}^{+})^*\right]$ \\ 
$C_5$ & $16\left[ \absq{G_{AA}^{-} - G_{VV}^{+}} + \absq{G_{VA}^{+} - G_{AV}^{-}}\right] + 128 \absq{G_{TT}^{+}} +32 \mathrm{Re}\left[\left(G_{SS}^{-}+G_{PP}^{-}\right)(G_{TT}^{+})^*\right]$ \\ 
$C_6$ & $8\left[\absq{G_{PP}^{-}} + \absq{G_{SP}^{-}} + \absq{G_{PS}^{-}} + \absq{G_{SS}^{-}}\right] - 64 \absq{G_{TT}^{+}}$ \\ \hline
$C_7$ & $16\mathrm{Re}\left[ G_{PP}^{-} (G_{SP}^{-})^* + G_{SS}^{-} (G_{PS}^{-})^*\right]$ \\ 
$C_8$ & $16\mathrm{Re}\left[ G_{PP}^{-} (G_{SP}^{-})^* - G_{SS}^{-} (G_{PS}^{-})^*\right] - 32\mathrm{Re}\left[ G_{AA}^{-} (G_{VA}^{+})^* - G_{VV}^{+} (G_{AV}^{-})^*\right]$ \\ 
$C_9$ & $32\mathrm{Re}\left[\left(G_{VV}^{+}+G_{AA}^{-}\right)\left((G_{AV}^{-})^*+(G_{VA}^{+})^*\right)\right] - 32\mathrm{Re}\left[\left(G_{SP}^{-}+G_{PS}^{-}\right)(G_{TT}^{+})^*\right]$ \\ 
$C_{10}$ & $32\mathrm{Re}\left[\left(G_{VV}^{+}-G_{AA}^{-}\right)\left((G_{AV}^{-})^*-(G_{VA}^{+})^*\right)\right] + 32\mathrm{Re}\left[\left(G_{SP}^{-}+G_{PS}^{-}\right)(G_{TT}^{+})^*\right]$ \\ 
$C_{11}$ & $16\mathrm{Re}\left[G_{AA}^{-} (G_{SP}^{-})^* + G_{AV}^{-} (G_{SS}^{-})^* - G_{VA}^{+} (G_{PP}^{-})^* - G_{VV}^{+} (G_{PS}^{-})^*\right] - 32 \mathrm{Re}\left[\left(G_{AV}^{-}-G_{VA}^{+}\right)(G_{TT}^{+})^*\right]$ \\ 
$C_{12}$ & $16\mathrm{Re}\left[G_{AA}^{-} (G_{SP}^{-})^* - G_{AV}^{-} (G_{SS}^{-})^* - G_{VA}^{+} (G_{PP}^{-})^* + G_{VV}^{+} (G_{PS}^{-})^*\right] - 32 \mathrm{Re}\left[\left(G_{AV}^{-}+G_{VA}^{+}\right)(G_{TT}^{+})^*\right]$ \\ 
$C_{13}$ & $-32 m_N\mathrm{Im}\left[\left(G_{PP}^{-}+G_{SS}^{-}\right)(G_{TT}^{+})^*\right]$ \\ 
& $+16m_\alpha \left( \mathrm{Im}\left[ G_{PP}^{-} (G_{AA}^{-})^* + G_{PS}^{-} (G_{AV}^{-})^* - G_{SP}^{-} (G_{VA}^{+})^* - G_{SS}^{-}(G_{VV}^{+})^*\right]  -2 \mathrm{Im}\left[\left(G_{VV}^{+}-G_{AA}^{-}\right)(G_{TT}^{+})^*\right]\right)$ \\
& $+16m_\beta \left( \mathrm{Im}\left[ -G_{PP}^{-} (G_{AA}^{-})^* + G_{PS}^{-} (G_{AV}^{-})^* + G_{SP}^{-} (G_{VA}^{+})^* - G_{SS}^{-}(G_{VV}^{+})^*\right] + 2\mathrm{Im}\left[\left(G_{AA}^{-}+G_{VV}^{+}\right)(G_{TT}^{+})^*\right]\right)$ \\\hline\hline
\end{tabular}
\end{center}
\end{table}

\begin{table}[!h]
\begin{center}
\caption{Identical to Table~\ref{table:MSqMajGen}, for the decays of a MF $N$, but now assuming only neutral mediators contribute to the decay. Here we have factored out a common $m_\phi^{-4}$ from each coefficient.\label{table:MSqMajNMO}}
\small
\begin{tabular}{c|c}\hline\hline
$C_i$ & Majorana Fermion $N$, Neutral Mediators Only  \\ \hline
$C_1$ & $32\left(g_{\nu P}^2 + g_{\nu S}^2\right) \left( \absq{g_{\ell P}} - \absq{g_{\ell S}}\right) + 64\left(g_{\nu A}^2 + g_{\nu V}^2\right)\left( \absq{g_{\ell V}} - \absq{g_{\ell A}}\right)$ \\
$C_2$ & $384 g_{\nu T} \left(g_{\nu A} \mathrm{Im}\left( g_{\ell A} g_{\ell T}^*\right) + g_{\nu V} \mathrm{Re}\left(g_{\ell V} g_{\ell T}^*\right)\right) - 64\left(g_{\nu A}g_{\nu P} + g_{\nu S} g_{\nu V}\right) \left( \mathrm{Im}\left( g_{\ell P} g_{\ell A}^* + g_{\ell S} g_{\ell V}^*\right)\right)$ \\
$C_3$ & $384 g_{\nu T} \left(-g_{\nu A} \mathrm{Im}\left( g_{\ell A} g_{\ell T}^*\right) + g_{\nu V} \mathrm{Re}\left(g_{\ell V} g_{\ell T}^*\right)\right) - 64\left(g_{\nu A}g_{\nu P} + g_{\nu S} g_{\nu V}\right) \left( \mathrm{Im}\left(g_{\ell P} g_{\ell A}^* - g_{\ell S} g_{\ell V}^* \right)\right)$ \\
$C_4$ & $64\left(g_{\nu A}^2 + g_{\nu V}^2\right) \left( \absq{g_{\ell V}} + \absq{g_{\ell A}}\right) + 512 g_{\nu T}^2 \absq{g_{\ell T}} -128 g_{\nu T}\left( g_{\nu P}\mathrm{Re}\left(g_{\ell P} g_{\ell T}^*\right) - g_{\nu S}\mathrm{Im}\left(g_{\ell S}g_{\ell T}^*\right)\right)$ \\
$C_5$ & $64\left(g_{\nu A}^2 + g_{\nu V}^2\right) \left( \absq{g_{\ell V}} + \absq{g_{\ell A}}\right) + 512 g_{\nu T}^2 \absq{g_{\ell T}} +128 g_{\nu T}\left( g_{\nu P}\mathrm{Re}\left(g_{\ell P} g_{\ell T}^*\right) - g_{\nu S}\mathrm{Im}\left(g_{\ell S}g_{\ell T}^*\right)\right)$ \\
$C_6$ & $32\left(g_{\nu P}^2 + g_{\nu S}^2\right) \left( \absq{g_{\ell P}} + \absq{g_{\ell S}}\right) - 256 g_{\nu T}^2 \absq{g_{\ell T}}$ \\ \hline
$C_7 = C_8$ & $0$ \\
$C_9 = -C_{10}$ & $128\left(g_{\nu A}^2 + g_{\nu V}^2\right) \mathrm{Re}\left(g_{\ell V} g_{\ell A}^*\right) - 128 g_{\nu T}\left( -g_{\nu S}\mathrm{Im}\left(g_{\ell P} g_{\ell T}^*\right) + g_{\nu P}\mathrm{Re}\left(g_{\ell S}g_{\ell T}^*\right)\right)$ \\
$C_{11}$ & $64\left(g_{\nu A} g_{\nu S} - g_{\nu P} g_{\nu V}\right) \mathrm{Re}\left(g_{\ell P} g_{\ell A}^* + g_{\ell S} g_{\ell V}^*\right) + 128 g_{\nu T}\left( g_{\nu V}\mathrm{Re}\left(g_{\ell A} g_{\ell T}^*\right) + g_{\nu A} \mathrm{Im}\left(g_{\ell V} g_{\ell T}^*\right) \right)$ \\
$C_{12}$ & $64\left(g_{\nu A} g_{\nu S} - g_{\nu P} g_{\nu V}\right) \mathrm{Re}\left(g_{\ell P} g_{\ell A}^* - g_{\ell S} g_{\ell V}^*\right) - 128 g_{\nu T}\left( g_{\nu V}\mathrm{Re}\left(g_{\ell A} g_{\ell T}^*\right) - g_{\nu A} \mathrm{Im}\left(g_{\ell V} g_{\ell T}^*\right) \right)$ \\
$C_{13}$ & $-128 m_N g_{\nu T} \left(g_{\nu P}\mathrm{Im}\left(g_{\ell P}g_{\ell T}^*\right) + g_{\nu S}\mathrm{Re}\left(g_{\ell S}g_{\ell T}^*\right)\right)$ \\
 & $-64m_\alpha \left(g_{\nu A}g_{\nu P} + g_{\nu S} g_{\nu V}\right) \mathrm{Re}\left(g_{\ell P} g_{\ell A}^* + g_{\ell S} g_{\ell V}^*\right) + 128 m_\alpha g_{\nu T}\left(g_{\nu A} \mathrm{Re}\left(g_{\ell A} g_{\ell T}^*\right) - g_{\nu V}\mathrm{Im}\left(g_{\ell V} g_{\ell T}^*\right) \right)$ \\
 & $+64m_\beta \left(g_{\nu A}g_{\nu P} + g_{\nu S} g_{\nu V}\right) \mathrm{Re}\left(g_{\ell P} g_{\ell A}^* - g_{\ell S} g_{\ell V}^*\right) + 128 m_\beta g_{\nu T}\left(g_{\nu A} \mathrm{Re}\left(g_{\ell A} g_{\ell T}^*\right) + g_{\nu V}\mathrm{Im}\left(g_{\ell V} g_{\ell T}^*\right) \right)$ \\ \hline\hline
\end{tabular}
\end{center}
\end{table}

\begin{table}[!h]
\begin{center}
\caption{Identical to Table~\ref{table:MSqMajGen}, for a MF $N$, further assuming that the final-state charged leptons are identical.\label{table:MSqMajAB}}
\begin{tabular}{c|c}\hline\hline
$C_i$ & Majorana Fermion $N$ Decaying to Identical Final-State Charged Leptons \\ \hline
$C_1$ & $32\left[(I_{PP}^{-})^{2} + (R_{SP}^{-})^{2} - (I_{SS}^{-})^{2} - (R_{PS}^{-})^{2}\right] + 64\left[(R_{VV}^{+})^{2} + (I_{AV}^{-})^{2} - (I_{AA}^{-})^{2} - (R_{VA}^{+})^{2}\right]$ \\ 
$C_2$ & $64\left[R_{SP}^- R_{VA}^+ - I_{AA}^- I_{PP}^-\right] + 384 R_{VV}^+ R_{TT}^+$ \\ 
$C_3$ & 
$C_2$ \\ 
$C_4$ & $64\left[(R_{VV}^{+})^{2} + (I_{AV}^{-})^{2} + (I_{AA}^{-})^{2} + (R_{VA}^{+})^{2}\right] + 512 (R_{TT}^{+})^{2}$ \\ 
$C_5$ & 
$C_4$ \\ 
$C_6$ & $32\left[(I_{PP}^{-})^{2} + (R_{SP}^{-})^{2} + (I_{SS}^{-})^{2} + (R_{PS}^{-})^{2}\right] - 256 (R_{TT}^{+})^{2}$ \\ \hline 
$C_7$ & $0$ \\ 
$C_8$ & $0$ \\ 
$C_9$ & $128\left(I_{AA}^- I_{AV}^- + R_{VA}^+ R_{VV}^+ - R_{TT}^+ \left(R_{PS}^- + R_{SP}^-\right)\right)$ \\ 
$C_{10}$ & 
$-C_9$ \\ 
$C_{11}$ & $64\left(I_{AV}^- I_{SS}^- - R_{PS}^- R_{VV}^+ + 2 R_{TT}^+ R_{VA}^+\right)$ \\ 
$C_{12}$ & 
$-C_{11}$ \\ 
$C_{13}$ & $-128 m_N \left(I_{SS}^- + I_{PP}^-\right)R_{TT}^+ -128m_\ell \left( I_{AV}^- R_{PS}^- + I_{SS}^- R_{VV}^+ - 2 I_{AA}^- R_{TT}^+\right)$ \\ \hline\hline
\end{tabular}
\end{center}
\end{table}

\begin{table}[!h]
\begin{center}
\caption{Identical to Table~\ref{table:MSqMajAB}, a MF $N$ decaying to identical final-state charged leptons, but assuming only neutral mediators contribute to the decay. We have factored out a common $m_\phi^{-4}$ from each coefficient.\label{table:MSqMajABNMO}}
\small
\begin{tabular}{c|c}\hline\hline
$C_i$ & Majorana Fermion $N$, Identical Final-State Charged Leptons, Neutral Mediators Only  \\ \hline
$C_1$ & $128 \left(g_{\ell P}^2 - g_{\ell S}^2\right) \left(g_{\nu P}^2 + g_{\nu S}^2\right) + 256 \left( g_{\ell V}^2 - g_{\ell A}^2\right) \left( g_{\nu V}^2 + g_{\nu A}^2\right)$ \\
$C_2 = C_3$ & $256 \left(6g_{\ell V} g_{\nu V} g_{T} - g_{\ell A} g_{\ell P}\left(g_{\nu A} g_{\nu P} + g_{\nu S} g_{\nu V}\right)\right)$ \\
$C_4 = C_5$ & $256\left( \left(g_{\ell A}^2 + g_{\ell V}^2\right)\left(g_{\nu A}^2 + g_{\nu V}^2\right) + 8g_{T}^2\right)$ \\
$C_6$ & $128 \left( \left(g_{\ell P}^2 + g_{\ell S}^2\right)\left(g_{\nu P}^2 + g_{\nu S}^2\right) - 8g_{T}^2\right)$ \\ \hline
$C_7 = C_8$ & $0$ \\
$C_9 = -C_{10}$ & $512 \left( g_{\ell A} g_{\ell V} \left(g_{\nu V}^2 + g_{\nu A}^2\right) + g_{T} \left(g_{\ell P} g_{\nu S} - g_{\ell S} g_{\nu P}\right) \right)$ \\
$C_{11} = -C_{12}$ & $256 \left(g_{\ell S} g_{\ell V} \left(g_{\nu A} g_{\nu S} - g_{\nu P} g_{\nu V}\right) + 2g_{T} g_{\ell A} g_{\nu V}\right)$ \\
$C_{13}$ & $-512 m_N g_{T} \left(g_{\ell P} g_{\nu P} + g_{\ell S} g_{\nu S}\right)$ \\
 & $+ 512 m_{\ell} \left( 2g_{T} g_{\ell A} g_{\nu A} - g_{\ell S} g_{\ell V} \left(g_{\nu A} g_{\nu P} + g_{\nu S} g_{\nu V}\right) \right)$ \\ \hline \hline
 \end{tabular}
\end{center}
\end{table}

\newpage
\section{Spin-dependence in a Charge-Blind Detector}
Above, we calculated the matrix-elements squared for the decays of a Dirac fermion $N$ for $N \to \nu \ell_\alpha^+ \ell_\beta^-$ and $\overline{N} \to \overline{\nu} \ell_\alpha^+ \ell_\beta^-$, where we implicitly assumed that we were in a situation in which the charge and particle identification of the final-state charged leptons was possible. Let us now imagine that we are in a scenario in which the particle identity of the leptons is easy (for instance, discriminating electrons from muons in a liquid/gaseous argon time projection chamber), but measuring their charge is impossible, or at least difficult (for instance, in a detector that is not magnetized). 

As an example, we focus on the case where $N$ couples only to the Standard Model muon-flavored lepton doublet. In the typical scenario where the HNL interactions are only via mixing with the light neutrinos, this implies that $U_{\mu N}$ is the only nonzero mixing angle present. If $N$ is a Dirac fermion, then $N$ (with lepton number 1) can only decay to the final state $\nu \mu^- e^+$, but not the state $\nu \mu^+ e^-$. Likewise, its counterpart $\overline{N}$ (with lepton number $-1$) can only decay to $\overline{\nu} \mu^+ e^-$, but not $\overline{\nu} \mu^- e^+$. Given the above calculations, it is possible that, even if $N$ is a Majorana fermion, that searches for the final state $\nu \mu^- e^+$ can be anisotropic, since its contributions are only from a matrix element that looks like $\mathcal{M}_1$ from Eq.~\eqref{eq:MN} and not those like $\mathcal{M}_2$ from Eq.~\eqref{eq:MNBar}.

However, if $N$ is a Majorana fermion, it will decay into the final states $\nu \mu^- e^+$ and $\nu \mu^+ e^-$ with equal likelihood (assuming CP invariance), and, if our detection technique is insensitive to the differences (i.e. the charges) in these final states, we must sum them incoherently before asking whether the final resulting distribution is isotropic or not. In general, given the above framework, the decays for a Majorana fermion $N$ into one specific-charge final state are given by the matrix-element-squared $\left\lvert \mathcal{M}_1 + \mathcal{M}_2\right\rvert^2$, whereas the decays into the opposite-charge final state are given by the matrix-element-squared $\left\lvert \mathcal{M}_1^c + \mathcal{M}_2^c\right\rvert^2$. Again, by assuming CP invariance, these must give the same total rate.

If our detector is insensitive to the charge of the final-state leptons (i.e. in a detector with no magnetic field), then, before determining any observables regarding isotropy, we must sum incoherently over these two final states\footnote{Some care must be taken in this scenario. In our general analyses, we considered contributions to the matrix element squared for different Lorentz-invariant contributions of final-state momentum four vectors. We labelled those four-vectors according to which charge of lepton they correspond to, i.e. $p_m^\sigma$ ($p_p^\sigma$) for the negatively-(positively-)charged lepton. Now, if charge is not measurable, we need to label the final-state charged leptons according to their flavor, e.g. $p_e^\sigma$ and $p_\mu^\sigma$. When this relabelling is performed, the cancellations discussed here follow.}, obtaining a decay distribution that will be proportional to $\left\lvert \mathcal{M}_1 + \mathcal{M}_2\right\rvert^2 + \left\lvert \mathcal{M}_1^c + \mathcal{M}_2^c\right\rvert^2$. We find that, regardless of the model assumptions made (neutral mediators, etc.), when we perform this calculation in the most general way possible and perform this incoherent sum, then all of the coefficients of spin-dependent Lorentz Invariants, $C_j$ for $7 \leq j \leq 13$ vanish completely: if a detector is charge blind and $N$ is a Majorana fermion, all sensitivity to the spin-dependence completely vanishes. This means that, in this case, the decays $N \to \nu \ell_\alpha^\pm \ell_\beta^\mp$ can be treated as (a) occurring with equal rate for the two final-state charges and (b) isotropic in the rest frame of $N$.

\section{Lorentz Invariants in $N$ Rest Frame
}\label{app:KinIntegration}
Here, we express the $13$ Lorentz invariant quantities $K_i$ in terms of the phase space parameters $m_{\ell\ell}^2$ (the invariant mass-squared of the charged lepton pair), $m_{\nu m}^2$ (the invariant mass-squared of the neutrino and the negatively charged lepton), $\cos\theta_{\ell\ell}$, $\gamma_{\ell\ell}$, and $\phi$. 
Fig.~\ref{fig:Kinematics} defines the $N$ rest-frame kinematics. Since the differential decay width does not depend on the azimuthal angle $\phi$, for concreteness, we fix $\phi=\pi/2$ so the three-momentum of the neutrino (and the sum of the charged lepton momenta, $p_{\ell\ell}=p_\alpha+p_\beta$) are in the $y-z$ plane. In this case, the neutrino and negatively charged lepton have, respectively, four momenta
\begin{align}
p_\nu &= E_\nu \left(1,0,-\sin\theta_{\ell\ell},-\cos\theta_{\ell\ell} \right)~,\\
p_m^0 &= E_m, \\
p_m^x &= \sqrt{E_m^2 - m_m^2} \sin\theta_{m\nu} \sin\gamma_{\ell\ell}, \\
p_m^y &= -\sqrt{E_m^2 - m_m^2} \left( \cos\theta_{m\nu}\sin\theta_{\ell\ell} + \sin\theta_{m\nu} \cos\gamma_{\ell\ell}\cos\theta_{\ell\ell}\right), \\
p_m^z &= \sqrt{E_m^2 - m_m^2} \left(\sin\theta_{m\nu}\cos\gamma_{\ell\ell}\sin\theta_{\ell\ell} - \cos\theta_{m\nu}\cos\theta_{\ell\ell}\right),
\end{align}
with
\begin{align}
E_m & = \dfrac{m_{\ell\ell}^2 + m_{\nu m}^2 - m_p^2}{2m_N}~,\\
\cos\theta_{m\nu} &\equiv \dfrac{\vec{p}_m \cdot \vec{p}_\nu}{|\vec{p}_m| |\vec{p}_\nu|} = \frac{E_m (m_N^2 - m_{\ell\ell}^2) - m_N \left( m_{\nu m}^2 - m_m^2 \right)}{\left(m_N^2 - m_{\ell\ell}^2\right) \sqrt{E_m^2 - m_m^2}}~.
\end{align}
The opening angle between the outgoing negatively-charged lepton and the neutrino $\theta_{m\nu}$ is determined entirely by energetics.  The positively charged lepton four momentum is determined by conservation of energy and momentum.
Using these forms for the momenta, the spin-independent $K_i$ are
\begin{align}
K_1 &= m_m m_p (p_\nu p_N) = \frac{1}{2} m_m m_p \left( m_N^2 - m_{\ell \ell}^2 \right), \label{eq:appK1}\\
K_2 &= m_m m_N (p_p p_\nu) = \frac{1}{2} m_m m_N \left( m_N^2 + m_m^2 - m_{\nu m}^2 - m_{\ell \ell}^2 \right), \\
K_3 &= m_p m_N (p_m p_\nu) = \frac{1}{2} m_p m_N \left( m_{\nu m}^2 - m_m^2 \right), \\
K_4 &= (p_p p_N)(p_m p_\nu) = \frac{1}{4} \left( m_{\nu m}^2 - m_m^2 \right) \left( m_N^2 + m_p^2 - m_{\nu m}^2  \right), \\
K_5 &= (p_m p_N)(p_p p_\nu) = \frac{1}{4} \left( m_{\ell\ell}^2 + m_{\nu m}^2 - m_p^2 \right) \left( m_N^2 + m_m^2 - m_{\ell\ell}^2 - m_{\nu m}^2 \right), \\
K_6 &= (p_p p_m)(p_\nu p_N) = \frac{1}{4} \left( m_N^2 - m_{\ell \ell}^2 \right) \left(m_{\ell\ell}^2 - m_m^2 - m_p^2 \right).
\end{align}
The spin-dependent $K_i$ are
\begin{align}
K_7 &= m_N (p_p p_m) (p_\nu s) = \frac{1}{4} \left( m_N^2 - m_{\ell\ell}^2 \right)\left( m_{\ell\ell}^2 - m_p^2 - m_m^2 \right) \cos\theta_{\ell\ell}~,\\
K_8 &= m_N m_m m_p (p_\nu s) = \frac{1}{2} m_m m_p \left(m_N^2 - m_{\ell\ell}^2\right)  \cos\theta_{\ell\ell} ~,\\
K_9 &= m_N (p_m p_\nu)(p_p s) \nonumber \\ 
&= \frac{m_N}{2}\left( m_{\nu m}^2 -m_m^2 \right)\left( |\vec{p}_m|\left(\cos\gamma_{\ell\ell} \sin\theta_{\ell\ell} \sin\theta_{m\nu} - \cos\theta_{\ell\ell} \cos\theta_{m\nu} \right) - \frac{m_N^2 - m_{\ell\ell}^2}{2m_N}\cos\theta_{\ell\ell} \right)~,\\
K_{10} &= m_N (p_p p_\nu) (p_m s)  \nonumber \\
&= -\frac{m_N}{2}\left(m_N^2 + m_m^2 - m_{\nu m}^2 -m_{\ell\ell}^2 \right)|\vec{p}_m|\left(\cos\gamma_{\ell\ell} \sin\theta_{\ell\ell} \sin\theta_{m\nu} - \cos\theta_{\ell\ell} \cos\theta_{m\nu} \right)~,\\
K_{11} &= m_m \left( (p_\nu p_N)(p_p s) - (p_p p_N)(p_\nu s)\right) \nonumber \\
&= \frac{m_m}{2}  \left( m_N^2 - m_{\ell\ell}^2 \right) \left[ |\vec{p}_m| \left( \cos\gamma_{\ell\ell} \sin\theta_{\ell\ell}  \sin\theta_{m\nu} - \cos\theta_{\ell\ell} \cos\theta_{m\nu} \right) -  \frac{2m_N^2 + m_p^2 - m_{\ell\ell}^2 - m_{\nu m}^2}{2 m_N}\cos\theta_{\ell\ell}\right]~,\\
K_{12} &= m_p \left( (p_\nu p_N)(p_m s) - (p_m p_N)(p_\nu s)\right) \nonumber \\
&= \frac{m_p}{2}  \left( m_N^2 - m_{\ell\ell}^2 \right) \left[ - |\vec{p}_m| \left( \cos\gamma_{\ell\ell} \sin\theta_{\ell\ell}  \sin\theta_{m\nu} - \cos\theta_{\ell\ell} \cos\theta_{m\nu} \right) + \frac{ m_p^2 - m_{\nu m}^2 - m_{\ell\ell}^2 }{2 m_N}\cos\theta_{\ell\ell} \right]~,\\
K_{13} &= \varepsilon_{\rho\sigma\lambda\eta} p_m^\rho p_p^\sigma p_\nu^\lambda s^\eta = \frac{1}{2}(m_N^2 - m_{\ell\ell}^2) |\vec{p}_m| \sin\theta_{\ell\ell} \sin\gamma_{\ell\ell} \sin\theta_{m\nu}~.
\label{eq:appK13}
\end{align}

\section{Integration over Invariant Masses and Angular Dependence of Distribution}\label{app:Integration}
In this appendix, we demonstrate how the differential partial width introduced in Eq.~\eqref{eq:dGamma}, a linear combination of the $K_i$ explored in Appendix~\ref{app:KinIntegration}, can be integrated with respect to the invariant masses $m_{\ell\ell}^2$ and $m_{\nu m}^2$ (as well as the unphysical angle $\phi$). This allows us to obtain the differential partial width depending only on the angles $\cos\theta_{\ell\ell}$ and $\gamma_{\ell\ell}$, with which we can discuss the (an)isotropy of $N$ decay and the forward/backward asymmetry $A_{\rm FB}$ for this process.

\begin{table}[htbp]
\begin{center}
\caption{Weights of the spin-independent Lorentz Invariants for their contributions to the double-differential partial width as written in Eq.~(\ref{eq:TripleDiff}). \label{Tab:LIContributionsSI}}
\begin{tabular}{c|cccc}\hline\hline
Lorentz Invariant & $I_0^i$ & $I_1^i$ & $I_2^i$ & $I_3^i$ \\ \hline
$K_1 = m_m m_p (p_\nu p_N)$ & 0 & $\sigma^2 - \delta^2$ & 0 & 0 \\ 
$K_2 = m_m m_N (p_p p_\nu)$ & 0 & $\sigma+\delta$ & $-\sigma\delta(\sigma+\delta)$ & 0 \\ 
$K_3 = m_p m_N (p_m p_\nu)$ & 0 & $\sigma-\delta$ & $\sigma\delta(\sigma-\delta)$ & 0 \\
$K_4 = (p_p p_N)(p_m p_\nu)$ & $\frac{2}{3}$ & $\frac{1}{6}(2-\sigma^2-\delta^2)$ & $\frac{1}{6}\left(\sigma^2 + \delta^2 - 2\sigma^2\delta^2\right)$ & $-\frac{2}{3} \sigma^2 \delta^2$ \\ 
$K_5 = (p_m p_N)(p_p p_\nu)$ & $\frac{2}{3}$ & $\frac{1}{6}(2-\sigma^2-\delta^2)$ & $\frac{1}{6}\left(\sigma^2 + \delta^2 - 2\sigma^2\delta^2\right)$ & $-\frac{2}{3} \sigma^2 \delta^2$ \\ 
$K_6 = (p_p p_m)(p_\nu p_N)$ & 2 & $-(\sigma^2 + \delta^2)$ & 0 & 0 \\ \hline\hline
\end{tabular}
\end{center}
\end{table}
It is useful to introduce dimensionless variables 
\begin{eqnarray}
z_{\ell\ell} & \equiv \dfrac{m_{\ell\ell}^2}{m_N^2}~,& \sigma  \equiv \frac{m_m + m_p}{m_N}~,  \label{eq:sigma} \\
z_{\nu m} & \equiv \dfrac{m_{\nu m}^2}{m_N^2}~,&  \delta  \equiv \frac{m_m - m_p}{m_N}~.  \label{eq:delta}
\end{eqnarray}
The region of accessible phase space is the usual Dalitz region \cite{PDG2}, with $\sigma^2 \le z_{\ell\ell}\le 1$ and the minimum/maximum of $z_{\nu m}$ given by
\begin{equation}
z_{\nu m}^{\rm min./max.} = \frac{1}{4}\left( \sigma^2 +\delta^2 + 2\frac{\sigma\delta}{z_{\ell\ell}} + 2(1-z_{\ell\ell}) \right) \mp \frac{1-z_{\ell\ell}}{2z_{\ell\ell}}\sqrt{\left( z_{\ell\ell} - \delta^2 \right) \left( z_{\ell\ell} - \sigma^2 \right)}
\end{equation}
Upon integrating the differential distribution over $z_{\nu m}$ and $\phi$, the distribution is linear in $\cos\theta_{\ell\ell}$. We express this quantity using
\begin{align}\label{eq:TripleDiff}
\frac{d\Gamma}{d\cos\theta_{\ell\ell}\, d\gamma_{\ell\ell}\, d z_{\ell\ell}} &= \frac{m_N^5}{2^{13}\pi^4} \left(1-z_{\ell\ell}\right)^2 \sqrt{z_{\ell\ell} - \sigma^2}\sqrt{z_{\ell\ell} - \delta^2} \sum_{m=0}^{3} z_{\ell\ell}^{-m}\left(
\sum_{i=1}^{6} C^i I_m^i + \sum_{i=7}^{13} C^i D_m^i\cos\theta_{\ell\ell} \right) \\ \nonumber
&+\frac{m_N^5}{2^{15}\pi^3}\sin\theta_{\ell\ell}\frac{(1 - z_{\ell\ell})^2 (z_{\ell\ell} - \sigma^2)(z_{\ell\ell} - \delta^2)}{z_{\ell\ell}^{5/2}} \sum_{i=9}^{13}\left(C^i \kappa_{C}^{i}\cos\gamma_{\ell\ell} + C^{i} \kappa_{S}^{i}\sin\gamma_{\ell\ell}\right).
\end{align}
The $C^i$ are the coefficients for each Lorentz Invariant that enter the matrix-element-squared, given in Tables~\ref{table:MSqDirac}-\ref{table:MSqMajAB}.
The spin-independent factors $I_m^i$ and the spin-dependent factors $D_m^i$ are both functions of $\delta$ and $\sigma$ only, and are given in Tables~\ref{Tab:LIContributionsSI} and~\ref{Tab:LIContributionsSD}. The $C_i$ only contribute to the total width for the spin-independent factors $I_m^i$ for $1 \leq i \leq 6$ and to the forward/backward asymmetry for the spin-dependent factors $D_m^i$ for $7 \leq i \leq 12$. The $\kappa_{C,S}^{i}$ terms appear proportional to either $\sin\theta_{\ell\ell}\cos\gamma_{\ell\ell}$ or $\sin\theta_{\ell\ell}\sin\gamma_{\ell\ell}$ and are nonzero only for $9 \leq i \leq 13$ -- their values are given in Table~\ref{Tab:KappaCS}.
\begin{table}[htbp]
\begin{center}
\caption{Weights of the spin-dependent Lorentz Invariants for their contributions to the double-differential partial width as written in Eq.~(\ref{eq:TripleDiff}). \label{Tab:LIContributionsSD}}
\begin{tabular}{c|cccc}\hline \hline
Lorentz Invariant & $D_0^i$ & $D_1^i$ & $D_2^i$ & $D_3^i$ \\ \hline
$K_7 = m_N (p_p p_m)(p_\nu s)$ & $2$ & $-\left( \sigma^2+\delta^2 \right)$ & 0 & 0 \\ 
$K_8 = m_N m_m m_p (p_\nu s)$ & 0 & $\left(\sigma^2-\delta^2\right)$ & 0 & 0 \\ 
$K_9 = m_N (p_m p_\nu)(p_p s)$ & $\frac{2}{3}$ & $-\frac{1}{6}(2 + \sigma^2 + \delta^2)$ & $-\frac{1}{6}\left(\sigma^2 + \delta^2 + 2\sigma^2\delta^2\right)$ & $\frac{2}{3}\sigma^2\delta^2$ \\ 
$K_{10} = m_N (p_p p_\nu) (p_m s)$ & $\frac{2}{3}$ & $-\frac{1}{6}(2 + \sigma^2 + \delta^2)$ & $-\frac{1}{6}\left(\sigma^2 + \delta^2 + 2\sigma^2\delta^2\right)$ & $\frac{2}{3}\sigma^2\delta^2$ \\ 
$K_{11} = m_m \left( (p_\nu p_N)(p_p s) - (p_p p_N)(p_\nu s)\right)$ & 0 & $-\left( \sigma+\delta \right)$ & $\sigma\delta(\sigma+\delta)$ & 0 \\ 
$K_{12} = m_p \left( (p_\nu p_N)(p_m s) - (p_m p_N)(p_\nu s)\right)$ & 0 & $\delta-\sigma$ & $\sigma\delta(\delta-\sigma)$ & 0 \\ \
$K_{13} = \varepsilon_{\rho\sigma\lambda\eta} p_m^\rho p_p^\sigma p_\nu^\lambda s^\eta$ & 0 & 0 & 0 & 0 \\ \hline \hline
\end{tabular}
\end{center}
\end{table}
\begin{table}[htbp]
\begin{center}
\caption{Terms $\kappa_i^{C}$ and $\kappa_i^{S}$ that enter the differential partial width with respect to $\cos\theta_{\ell\ell}$, $\gamma_{\ell\ell}$, and $z_{\ell\ell}$ in Eq.~\eqref{eq:TripleDiff}. All other $\kappa_i^{C,S}$ for the Lorentz Invariants $K_i$ not shown here are zero.\label{Tab:KappaCS}}
\begin{tabular}{c|cc}\hline\hline
Lorentz Invariant & $\kappa_i^C$ & $\kappa_i^S$ \\ \hline
$K_9 = m_N (p_m p_\nu)(p_p s)$ & $\sigma\delta + z_{\ell\ell}$ & 0 \\
$K_{10} = m_N (p_p p_\nu) (p_m s)$ & $\sigma\delta - z_{\ell\ell}$ & 0 \\
$K_{11} = m_m \left( (p_\nu p_N)(p_p s) - (p_p p_N)(p_\nu s)\right)$ & $\left(\sigma+\delta\right)z_{\ell\ell}$ & 0 \\
$K_{12} = m_p \left( (p_\nu p_N)(p_m s) - (p_m p_N)(p_\nu s)\right)$ & $\left(\delta - \sigma\right)z_{\ell\ell}$ & 0 \\
$K_{13} = \varepsilon_{\rho\sigma\lambda\eta} p_m^\rho p_p^\sigma p_\nu^\lambda s^\eta$ & 0 & $\dfrac{2z_{\ell\ell}}{m_N}$ \\ \hline\hline
\end{tabular}
\end{center}
\end{table}

Our next goal is to integrate Eq.~\eqref{eq:TripleDiff} over $z_{\ell\ell}$ in the range $\left[\sigma^2, 1\right]$. We express this double-differential partial with as
\begin{align}\label{eq:DoubleDiffAngle}
\frac{d\Gamma}{d\cos\theta_{\ell\ell}\, d\gamma_{\ell\ell}} &= \frac{m_N^5}{2^{13}\pi^4} \left[ \sum_{i = 1}^{6} C^i \left( \sum_{m=0}^{3} I_{m}^{i} T_m\right) + \sum_{i=7}^{12} C^i \left(\sum_{m=0}^{3} D_{m}^{i} T_m\right)\cos\theta_{\ell\ell}\right] \nonumber \\
&+ \frac{m_N^5 \sin\theta_{\ell\ell}}{2^{13}\cdot 3\cdot 5\cdot 7 \pi^3}\sum_{i=9}^{13} \left(C^i \widetilde{\kappa}_C^i \cos\gamma_{\ell\ell} + C^{i}\widetilde{\kappa}^{i}_S\sin\gamma_{\ell\ell}\right).
\end{align}
The quantities $T_m$ (which depend solely on $\sigma$ and $\delta$) contribute to the total width and forward/backward asymmetry of the decay and are relatively complicated functions. We give those separately in Appendix~\ref{eq:TIntegrals} so as not to disrupt the discussion here. The terms $\widetilde{\kappa}_{C,S}^{i}$ result from integrating the terms proportional to $\cos\gamma_{\ell\ell}$ and $\sin\gamma_{\ell\ell}$ in Eq.~\eqref{eq:TripleDiff}. These are
\begin{equation}
\left(\begin{array}{c}\widetilde{\kappa}_{C}^{9} \\ \widetilde{\kappa}_{C}^{10} \\ \widetilde{\kappa}_{C}^{11} \\ \widetilde{\kappa}_{C}^{12} \\ \widetilde{\kappa}_{S}^{13}\end{array}\right) = (1 - \sigma)^4 \times \left(\begin{array}{c} \left(4 + 16\sigma + 12\sigma^2 + 3\sigma^3\right) + 7\delta\sigma(4+\sigma) - 7\delta^2(4+\sigma) - 35\delta^3 \\ -\left(4 + 16\sigma + 12\sigma^2 + 3\sigma^3\right) + 7\delta\sigma(4+\sigma) + 7\delta^2(4+\sigma) - 35\delta^3 \\ \left(\sigma+\delta\right) \left( (4+16\sigma+12\sigma^2+3\sigma^3) - 7\delta^2(4+\sigma)\right) \\ \left(\delta-\sigma\right) \left( (4+16\sigma+12\sigma^2+3\sigma^3) - 7\delta^2(4+\sigma)\right) \\ \frac{2}{m_N} \left( (4+16\sigma+12\sigma^2+3\sigma^3) - 7\delta^2(4+\sigma)\right) \end{array}\right).
\end{equation}

In order to simplify Eq.~\eqref{eq:DoubleDiffAngle}, we define four quantities
\begin{align}
&X_{\Gamma} = \sum_{i=1}^{6} C^{i} \left( \sum_{m=0}^{3} I_m^{i} T_m\right), \\
&X_{\rm FB} = \sum_{i=7}^{12} C^{i} \left(\sum_{m=0}^{3} D_m^{i} T_m\right), \label{eq:XFBDef}\\
&X_{c_{\gamma}} = \frac{\pi}{105}\sum_{i=9}^{12} C^i \widetilde{\kappa}_i^C, \\
&X_{s_{\gamma}} =  \frac{\pi}{105} C^{13} \widetilde{\kappa}_{13}^S,
\end{align}
which allows us to write Eq.~\eqref{eq:DoubleDiffAngle} in a more compact form: 
\begin{equation}\label{eq:DoubleDiffAngleSimp}
\frac{d\Gamma}{d\cos\theta_{\ell\ell}\, d\gamma_{\ell\ell}} = \frac{1}{4\pi} \frac{m_N^5}{2^{11}\pi^3} \left[ X_{\Gamma} + X_{\rm FB}\cos\theta_{\ell\ell}  + \sin\theta_{\ell\ell} \left(X_{c_\gamma}\cos\gamma_{\ell\ell} + X_{s_\gamma} \sin\gamma_{\ell\ell}\right)\right].
\end{equation}

We can integrate this over either $\gamma_{\ell\ell}$ or $\cos\theta_{\ell\ell}$. Let us integrate $\gamma_{\ell\ell}$ first. We obtain
\begin{align}
\frac{d\Gamma}{d\cos\theta_{\ell\ell}} &= \frac{1}{2} \frac{m_N^5}{2^{11}\pi^3} X_{\Gamma} \left( 1 + \frac{X_{\rm FB}}{X_{\Gamma}} \cos\theta_{\ell\ell}\right), \\
&= \frac{\Gamma}{2} \left(1 + 2A_{\rm FB} \cos\theta_{\ell\ell}\right).
\end{align}
By definition,
\begin{align}
\Gamma &= \frac{m_N^5 X_{\Gamma}}{2^{11}\pi^3}, \\
A_{\rm FB} &\equiv \displaystyle\frac{\displaystyle\int_{0}^{1} \frac{d\Gamma}{d\cos\theta_{\ell\ell}} d\cos\theta_{\ell\ell} - \displaystyle\int_{-1}^{0} \frac{d\Gamma}{d\cos\theta_{\ell\ell}} d\cos\theta_{\ell\ell}}{\Gamma} = \frac{X_{\rm FB}}{2X_{\Gamma}}
\end{align}

If we had integrated Eq.~\eqref{eq:DoubleDiffAngleSimp} over $\cos\theta_{\ell\ell}$ instead, we obtain
\begin{align}
\frac{d\Gamma}{d\gamma_{\ell\ell}} &= \frac{1}{2\pi} \frac{m_N^5}{2^{11}\pi^3} \left[X_{\Gamma} + \frac{\pi}{4} \left(X_{c_\gamma}\cos\gamma_{\ell\ell} + X_{s_\gamma}\sin\gamma_{\ell\ell}\right)\right], \\
&= \frac{\Gamma}{2\pi} \left( 1 + \eta_C \cos\gamma_{\ell\ell} + \eta_S \sin\gamma_{\ell\ell}\right),
\end{align}
and by definition,
\begin{align}
\eta_{(C,S)} &= \frac{\pi X_{{(c,s)}_\gamma}}{4X_{\Gamma}}.
\end{align}

\section{Reference Integrals}\label{eq:TIntegrals}
As we found in Appendix~\ref{app:Integration}, determining the partial width of a certain channel, or the forward/backward asymmetry, requires integrating Eq.~\eqref{eq:DoubleDiffAngle} over $z_{\ell\ell}$. This amounts to determining the integrals
\begin{equation}
T_m \equiv \int_{\sigma^2}^1 \frac{(1-z)^2 \sqrt{z - \sigma^2} \sqrt{z - \delta^2}}{z^m} dz,
\end{equation}
where $0 \le \delta^2 \le \sigma^2 \le 1$ and $\sigma \ge 0$. The results for $m = 0,\ldots ,3$ are
\begin{align}
&\left( \begin{array}{c}T_0(\sigma,\delta) \\ T_1(\sigma,\delta) \\ T_2(\sigma,\delta) \\ T_3(\sigma,\delta)\end{array}\right) = \sqrt{1-\sigma^2}\sqrt{1-\delta^2} \left( \begin{array}{c} \frac{1}{192}\left(2 - \sigma^2 - \delta^2\right) \left( 8 - 8\sigma^2 - 8\delta^2 + 15\sigma^4 - 22\sigma^2 \delta^2 + 15\delta^4\right) \\
\frac{1}{24}\left( 8 + 10\sigma^2 + 10\delta^2 - 3\sigma^4 + 2\sigma^2\delta^2 - 3\delta^4\right)\\
-\frac{1}{4}\left(10 + \sigma^2 + \delta^2\right) \\
\frac{1}{4}\left(10 + \frac{1}{\sigma^2} + \frac{1}{\delta^2}\right) \end{array}\right) \nonumber \\
& +\log{\left(\frac{\sqrt{1-\sigma^2} + \sqrt{1-\delta^2}}{\sqrt{\sigma^2-\delta^2}}\right)} \left( \begin{array}{c} -\frac{1}{64}\left(\sigma^2-\delta^2\right)^2 \left(16 - 16\sigma^2 - 16\delta^2 + 5\sigma^4 + 6\sigma^2\delta^2 + 5\delta^4\right) \\
\frac{1}{8}\left( -8\sigma^2 - 8\delta^2 + 4\sigma^4 - 8\sigma^2\delta^2 + 4\delta^4 - \sigma^6 + \sigma^4\delta^2 + \sigma^2\delta^4 - \delta^6\right) \\
\frac{1}{4}\left(8 + 8\sigma^2 + 8 \delta^2 - \sigma^4 + 2\sigma^2\delta^2 - \delta^4\right) \\
-\left(4 + \sigma^2 + \delta^2\right)\end{array}\right) \nonumber \\
& + \log{\left(\frac{\sigma\sqrt{1-\delta^2} + \delta\sqrt{1-\sigma^2}}{\sigma\sqrt{1-\delta^2} - \delta\sqrt{1-\sigma^2}}\right)} \left(\begin{array}{c} 0 \\ \sigma\delta \\ -\frac{1}{2\sigma\delta}\left(\sigma^2 + \delta^2 + 4\sigma^2\delta^2\right) \\ \frac{1}{8\sigma^3\delta^3}\left( -\sigma^4 + 2\sigma^2\delta^2 - \delta^4 + 8\sigma^4\delta^2 + 8\sigma^2\delta^4 + 8\sigma^4\delta^4\right)\end{array}\right).
\end{align}
When $\delta = 0$, corresponding to the case of identical flavor final state leptons, these simplify to
\begin{align}
&\left( \begin{array}{c}T_0(\sigma,0) \\ T_1(\sigma,0) \\ T_2(\sigma,0) \\ T_3(\sigma,0)\end{array}\right) = \sqrt{1-\sigma^2}\left( \begin{array}{c} \frac{1}{192}\left(2 - \sigma^2\right) \left( 8 - 8\sigma^2 + 15\sigma^4\right) \\
\frac{1}{24}\left( 8 + 10\sigma^2 - 3\sigma^4\right)\\
-\frac{1}{4}\left(14 + \sigma^2\right) \\
\frac{1}{3}\left(13 + \frac{2}{\sigma^2}\right) \end{array}\right) \nonumber \\
& +\log{\left(\frac{1 + \sqrt{1-\sigma^2}}{\sigma}\right)} \left( \begin{array}{c} -\frac{1}{64}\sigma^4 \left(16 - 16\sigma^2 - 16\delta^2 + 5\sigma^4\right) \\
-\frac{1}{8}\sigma^2\left(8 - 4\sigma^2 + \sigma^4\right) \\
\frac{1}{4}\left(8 + 8\sigma^2 + - \sigma^4\right) \\
-\left(4 + \sigma^2\right)\end{array}\right).
\end{align}

\begin{figure}
\centering
\includegraphics[width=0.8\linewidth]{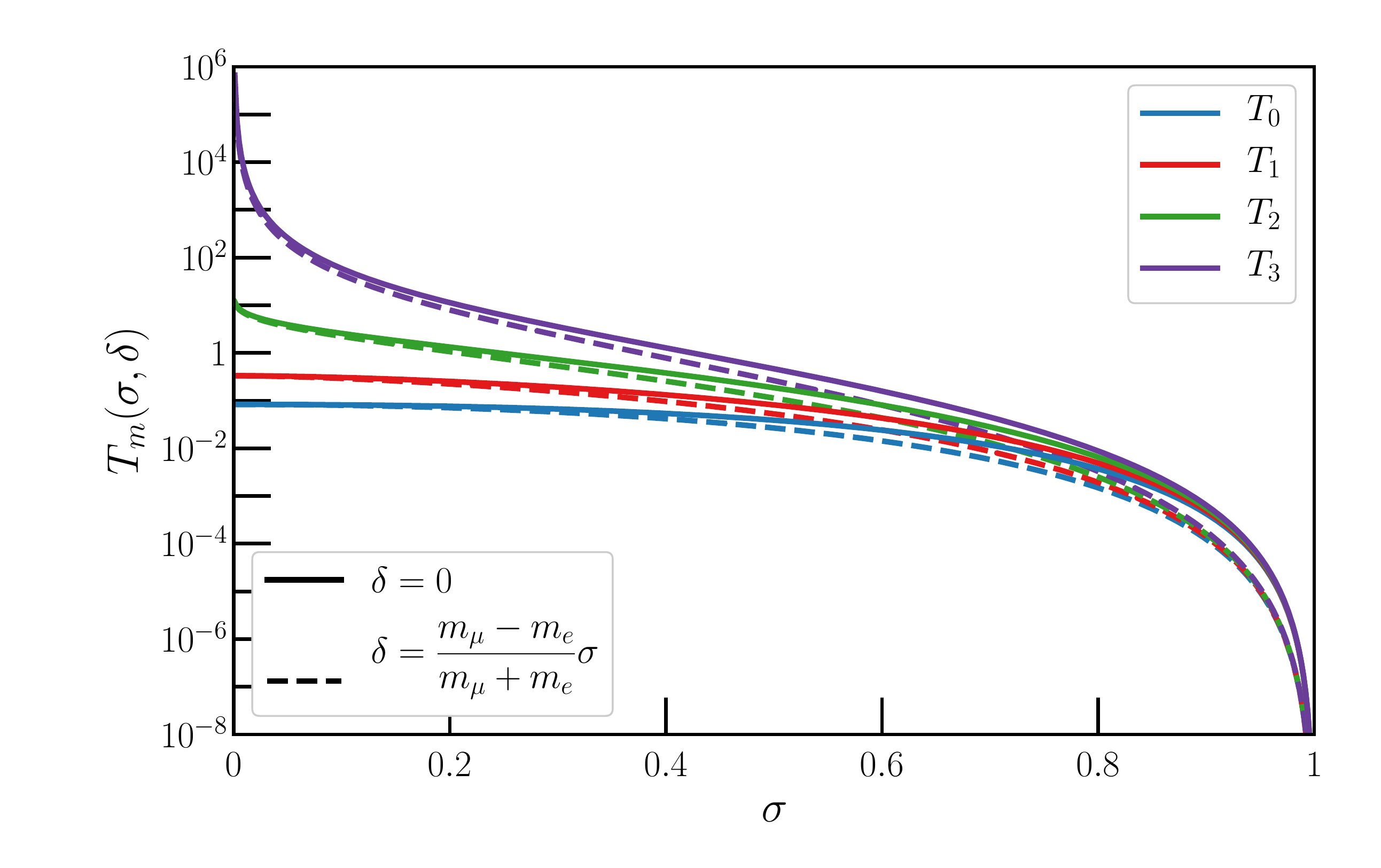}
\caption{Dependence of the four different $T_m(\sigma,\delta)$ as a function of $\sigma$ for two different choices of $\delta$: solid lines display $\delta = 0$, and dashed lines display $\delta/\sigma = (m_\mu - m_e)/(m_\mu + m_e) \approx 0.990$.\label{fig:Ts}}
\end{figure}

Fig.~\ref{fig:Ts} displays $T_m(\sigma, \delta)$ as a function of $\sigma$ for $m = 0$, $1$, $2$, $3$ for two choices of parameters. In solid lines, we plot $\delta = 0$ (corresponding to the results with identical final-state charged leptons. In dashed lines, we assume $\delta/\sigma = (m_\mu - m_e)/(m_\mu + m_e) \approx 0.990$, which would correspond to decays of the type $N \to \nu \mu^- e^+$.

\bibliographystyle{JHEP}
\bibliography{refs}

\end{document}